\documentclass[aps,prd,preprint,groupedaddress,showpacs]{revtex4}
\usepackage[dvips]{graphicx}

\begin{document}

\preprint{KOBE-TH-02-01}

\title{Stability of Fermi ball against deformation from spherical shape}

\author{T.Yoshida}
\affiliation{Department of Physics, Tokyo University, Hongo 7-3-1, Bunkyo-Ku, Tokyo 113-0033, Japan}

\author{K.Ogure}
\affiliation{Department of Physics, Kobe University, Rokkoudaicho 1-1, Nada-Ku, Kobe 667-8501, Japan}

\author{J.Arafune}
\affiliation{National Institution for Academic Degrees, Hitotsubashi 2-1-2, Chiyoda-Ku, Tokyo 101-8438, Japan}

\date{\today}

\begin{abstract}
The stability of a Fermi ball (F-ball), which is a kind of non-topological soliton accompanying the breakdown of the approximate $Z_2$ symmetry, is investigated in three situations: the case it is electrically neutral, the case it is electrically charged and unscreened, and the case it is electrically charged and screened.  We argue only the third case is physically meaningful since the neutral F-ball is unstable and the case of the unscreened charged one is observationally excluded when it has a sizable contribution to CDM.  We found that the energy scale of the breakdown of the approximate $Z_2$ symmetry $v$ should satisfy $v<3\times 10^6~\mbox{GeV}$ if the F-ball is a main component of CDM.  
\end{abstract}

\maketitle

\section{INTRODUCTION\label{introduction.sec}}

An F-ball, a kind of non-topological soliton, was introduced as a candidate for CDM \cite{Mac,Mor}. A similar object is also suggested to play an important role in baryogenesis \cite{Bra}.

\begin{figure}[htbp]
\includegraphics[height=5cm]{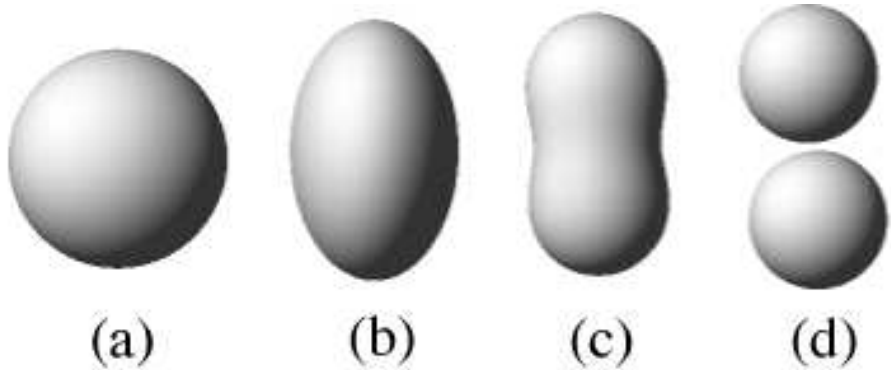}
\caption{A deformation and a fragmentation of the F-ball.  (a)
Sphere. (b) Cigar-like deformation.  (c)  Shape narrow in the middle.
(d) Fragmented F-balls.
\label{frag.fig}}
\end{figure}

The simplest type of the F-ball is a bubble of false vacuum surrounded by a thin domain wall, on which many zero-mode fermions are attached. The surface tension of the wall is balanced with the Fermi pressure due to the zero-mode fermions in the wall. The F-ball may be stabilized owing to this balance between the surface energy and the Fermi energy. Though the F-ball is stable against the variation of the radius in case it keeps a spherical shape, it may be unstable against deformation from sphere. Macpherson and Campbell pointed out in their pioneering work \cite{Mac} that such F-balls are unstable against deformation from the spherical shape, and that they should finally fragment into a number of tiny F-balls \footnote{This fragmentation is caused due to a small volume energy, which exists when the symmetry is biased \cite{Mac}.} (see Fig.\ref{frag.fig} for illustration of a fragmenting F-ball). 
Suppose there are a large number of such tiny F-balls in the present universe contributing sizably to CDM. The cosmic flux of these objects should be large, proportional to the inverse of the mass, and they should have been observed unless the scattering cross section with ordinary matter is very small. Here, we are not interested in such a new kind of weakly interacting particles since we need too many new assumptions in introducing new tiny particles and have only indirect means to detect such particles \cite{Mac}.

Morris then introduced a new idea, an electrically charged F-ball \footnote{When the fermions of the F-ball have electric U(1) charge, we call such an F-ball (electrically) charged F-ball hereafter. In the same manner, we define the terminology of the neutral F-ball for that Macpherson and Campbell originally introduced.} which is considered to be stabilized due to the repulsive long-range Coulomb force \cite{Mor}. Since the charged F-balls can be large and heavy in this case, they can sizably contribute to CDM without contradicting the present observations of the cosmic flux. These F-balls with a large cross section of interaction with ordinary matter are interesting because of their future detectability. (We nevertheless note that the cross section per unit mass for the F-ball $\sigma/M_f$ is not so large as the one which was recently proposed \cite{Spe, Tys} for the self-interacting dark matter. We do not go into details since the above issue is still controversial (see the discussions in Sec.\ref{conclusion.sec}).)

In the present paper, we first examine the instability of the neutral F-ball against the deformation from the spherical shape.  It is known \cite{Mac} that the spherical shape is unstable against deformation due to the finite volume energy within the thin-wall approximation.  We investigate the next-to-leading order approximation, the curvature effect of the wall, since it has a shape-dependent contribution to the total energy, and is important when the volume energy is very small. We find that the neutral F-ball is unstable even in the presence of the curvature effect.

We next examine whether the F-ball is stable when it is electrically charged and unscreened. Since the total energy of such an F-ball is higher than the sum of the energies of the fragmented F-balls, it fragments into smaller F-balls in finite time. Its lifetime is, however, longer than the age of the universe if there is an energy barrier high enough between the state of an F-ball and that of fragmented F-balls. We see that the Coulomb force does make the lifetime of the F-ball long enough.

We thirdly examine whether the F-ball has a lifetime long enough to survive until present if it is screened due to the electrons or positrons in the thermal bath of the early universe. Though the long ranged Coulomb force becomes short ranged in the thermal bath, the lifetime of the F-ball is found long enough in the proper region of the parameters.

We finally examine which kind of F-balls satisfy the conditions for a main component of CDM. We find that only the screened F-ball meets the conditions after taking into account the stability of the F-ball, observational constraints, and cosmological considerations.

The above contents are organized as follows: We first consider electrically neutral F-balls and discuss their stability in Sec.\ref{neutral.sec}. Small F-balls which are electrically charged and not screened by electrons are considered in Sec.\ref{unscreened.sec}.  Screened ones are considered in Sec.\ref{screened.sec}; this is the main theme in the present paper.  In Sec.\ref{constraint.sec}, we discuss certain constraints on the parameters obtained from various conditions.  We have summaries and discussions in Sec.\ref{conclusion.sec}.  The detailed calculations are given in the Appendices.

\section{STABILITY OF ELECTRICALLY NEUTRAL F-BALL\label{neutral.sec}}

\begin{figure}[htbp]
\includegraphics[height=5cm]{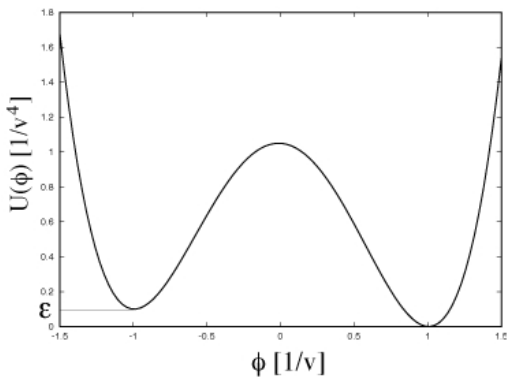}
\caption{The biased potential for $\lambda =1$. Two almost degenerate
vacua with the energy difference of $\epsilon$ exist: the true vacuum with $\phi=v$ and the false vacuum with $\phi=-v$.
\label{pot.fig}}
\end{figure}

We briefly introduce the electrically neutral F-ball proposed by Macpherson and Campbell \cite{Mac} to make clear the notations and technical terms. The Lagrangian density is given by
\begin{equation}
{\cal L}=\frac{1}{2}(\partial_{\mu}\phi)^2+\bar{\psi}_f(i\gamma^\mu\partial_{\mu}-G\phi)\psi_f-U(\phi)~,
\label{neu_lagrangian_def.eqn}
\end{equation}
where $\phi$ and $\psi_f$ are a scalar and a fermion field, respectively, and $G$ is a Yukawa coupling constant. Here, $U(\phi)$ is an approximate double-well potential  \footnote{Note that the qualitative discussions in the following do not depend on the explicit form of $U(\phi)$.},
\begin{equation}
U(\phi)=\frac{\lambda}{8}\left(\phi^2-v^2\right)^2+U_{\epsilon}(\phi)~.
\label{neu_phi4.eqn}
\end{equation}
The first term has the $Z_2$ symmetry under $\phi \leftrightarrow -\phi$, and the second term violates the symmetry, though we assume it is much smaller than the first one (see Fig.\ref{pot.fig}). 
Thus, the Lagrangian density has an approximate $Z_2$ symmetry, which we call 'biased $Z_2$ symmetry' \cite{Mac} hereafter.

After the phase transition breaks spontaneously the biased $Z_2$ symmetry at $T=T_{ph}\sim v$, there arise two almost degenerate vacua with the energy density difference of $\epsilon$. They are the true vacuum with $\phi=v$ and the false one with $\phi=-v$. Then a domain wall is produced between two vacua, and the fermions are captured as zero-modes in the domain wall \footnote{Such zero-mode solutions for $\psi_f$ can also exist in other cases, such as in the context of supersymmetry \cite{Mor2,Dva}.}. Though the effect of $\epsilon$ should be negligible soon after the phase transition, it gets more important as the universe expands. If the volume energy becomes larger than the surface energy during the expansion, the true vacuum is energetically favored and enlarges its volume. At the same time, the region of false vacuum gets diminished, and becomes a small confined region surrounded by the domain wall. Such a region of false vacuum would continue to shrink due to the surface tension and the volume effect proportional to $\epsilon$,  and finally disappear if we neglect the Fermi pressure caused by the zero-mode fermions trapped in the wall.  In our case, however, the Fermi pressure is not negligible and stops the shrinkage of the region, when it gets balanced with the surface tension. Such a bubble of the false vacuum with zero-mode fermions trapped in the surrounding wall is called F-ball \cite{Mac}.  We take the temperature for the F-ball production as $T=T_f$. Because $T_f$ is somewhat lower than $T_{ph}$, $T_f$ is assumed to be $T_f\lesssim 0.1v$ in the present paper.

\begin{figure}[htbp]
\includegraphics[height=6cm]{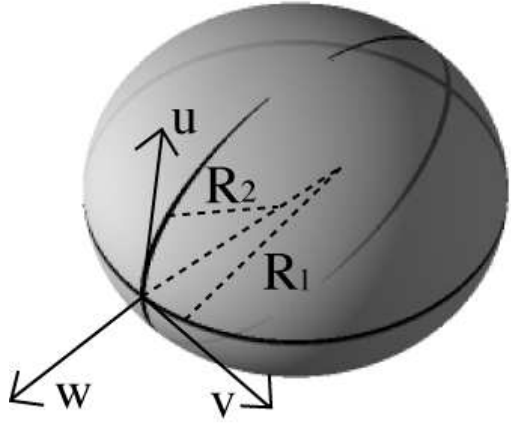}
\caption{The local orthogonal coordinates $(u,v,w)$, the origin of which is on the surface of the F-ball. We take $u$ and $v$ along the lines of curvature on the surface, and $w$ normal to the surface. Here, $R_1$ ($R_2$) is a radius of principal curvature with respect to $u=0$ ($v=0$) at the origin.
\label{localcoordinate.fig}}
\end{figure}

We consider a thin walled F-ball, the size of which is much larger than the thickness of the domain wall. For simplicity, we neglect the small energy difference $\epsilon$ and take the exact $Z_2$ symmetry until Sec.\ref{constraint.sec}. Then the Lagrangian density is
\begin{equation}
{\cal L}=\frac{1}{2}(\partial_{\mu}\phi)^2+\bar{\psi}_f(i\gamma^\mu\partial_{\mu}-G\phi)\psi_f-\frac{\lambda}{8}\left(\phi^2-v^2\right)^2~.
\label{neu_lagrangian.eqn}
\end{equation}
 The total energy of an F-ball is expressed as
\begin{equation}
E_{tot}=E_s+E_f~,
\label{neu_etot.eqn}
\end{equation}
where $E_s$ and $E_f$ are a contribution from the domain wall and that from the zero-mode fermions, respectively. In the thin wall case, Fermi gas in the wall distributes two dimensionally on the surface of the F-ball \footnote{In our context, we call the plane where the expectation value of $\phi$ vanishes "surface of the F-ball".}. Taking $w$-axis normal to the surface which is positive (negative) for the outside (inside) of the F-ball, we express the fermion number density as
\begin{equation}
n_f=\sigma_f\delta(w)~,
\label{neu_nfdef.eqn}
\end{equation}
with $\sigma_f$ the surface density of the fermion number, and the fermi energy,
\begin{equation}
E_f=\frac{4\sqrt{\pi}}{3}\int\mbox{d}S~\sigma_f^{3/2}~.
\label{neu_efdef.eqn}
\end{equation}
We obtain the surface energy,
\begin{equation}
E_s=\int\mbox{d}S\mbox{d}w~\left| (1+\frac{w}{R_1})(1+\frac{w}{R_2}) \right| \left\{ \frac{1}{2}\left(\nabla\phi\right)^2+\frac{\lambda}{8} \left( \phi^2-v^2 \right)^2 \right\}~,
\label{neu_esdef.eqn}
\end{equation}
where $R_1$ and $R_2$ are the radii of principal curvature of the F-ball surface (see Fig.\ref{localcoordinate.fig} and Appendix \ref{dv.sec}).

If the fermion-scalar coupling $G$ is not too small, the F-ball is stable against releasing a fermion with mass $\sim Gv$ from the wall to the vacuum.  The number of the fermions of the F-ball, 
\begin{equation}
N_f=\int\mbox{d}S~\sigma_f~,
\label{neu_nftot.eqn}
\end{equation}
is then conserved.  Thus, the energy of the F-ball is obtained by minimizing
\begin{equation}
E_{tot;\mu}=E_{tot}+\mu\left(N_f-\int\mbox{d}S~\sigma_f\right)~,
\label{neu_etotlambda.eqn}
\end{equation}
with $\mu$ the Lagrange multiplier. We first minimize it with respect to $\sigma_f$. Noting that $E_s$ is independent of $\sigma_f$, we obtain the uniform distribution of the fermions,
\begin{equation}
\sigma_f=\frac{N_f}{S}~,
\label{neu_uniform_distr.eqn}
\end{equation}
with $S$ the area of the F-ball surface. We next minimize the energy with respect to $\phi$. Assuming that the variation of $\phi$ in the direction parallel to the F-ball surface is much smaller than that in the direction $w$, we get the Euler-Lagrange equation for $\phi$,
\begin{equation}
\ddot{\phi}+\left( \frac{1}{R_1+w}+\frac{1}{R_2+w} \right)\dot{\phi}=\frac{\lambda}{2}\phi(\phi^2-v^2)~,
\label{neu_eqM.eqn}
\end{equation}
with the symbol 'dot' in $\dot{\phi}$ or $\ddot{\phi}$ standing for the derivative with respect to $w$. Here, $\phi$ fulfills the boundary condition,
\begin{eqnarray}
\phi = \left\{ 
\begin{array}{ll}
+v & (w\gg \delta_n) \\
0 & (w = 0) \\
-v & (w\ll -\delta_n)~~~,
\end{array}
\right.
\label{neu_bc.eqn}
\end{eqnarray}
with $\delta_n$ the width of the wall. We take the leading contribution, $\phi_0$, in the thin-wall expansion with $R_1,R_2\rightarrow\infty$. It satisfies
\begin{equation}
\ddot{\phi_0}=\frac{\lambda}{2}\phi_0(\phi_0^2-v^2)~,
\label{neu_eqM0.eqn}
\end{equation}
with
\begin{eqnarray}
\phi_0 = \left\{ 
\begin{array}{ll}
+v & (w\gg \delta_n) \\
0 & (w = 0) \\
-v & (w\ll -\delta_n)~~~.
\end{array}
\right.
\label{neu_bc0.eqn}
\end{eqnarray}
The solution for $\phi_0$,
\begin{equation}
\phi_0=v\tanh{\frac{w}{\delta_n}}~,
\label{neu_phi0.eqn}
\end{equation}
with $\delta_n=2/(\sqrt{\lambda}v)$, minimizes the energy under the condition, Eq.(\ref{neu_uniform_distr.eqn}), to be
\begin{equation}
E_{tot}^{(0)}=\Sigma S+\frac{4\sqrt{\pi}N_f^{3/2}}{3\sqrt{S}}~,
\label{neu_etot0.eqn}
\end{equation}
with $\Sigma=2\sqrt{\lambda}v^3/3$ the surface tension in the wall. We finally minimize $E_{tot}^{(0)}$ with respect to $S$, and obtain
\begin{equation}
E_{tot}^{(0)}=(12\pi\Sigma)^{1/3}N_f~,
\label{neu_etotmin0.eqn}
\end{equation}
with the area of the surface, $S=(2\sqrt{\pi}/3\Sigma)^{2/3}N_f$.  An F-ball can deform with this surface area being kept constant (see Fig.\ref{frag.fig}(a) to Fig.\ref{frag.fig}(c)). Moreover, it has the same energy even if it splits into some smaller F-balls (see Fig.\ref{frag.fig}(d)) since $E_{tot}^{(0)}$ is proportional to the number of fermions. Thus, we cannot tell whether the F-ball is stable or not against deformation and fragmentation into pieces, within the thin-wall approximation which corresponds to the leading order in the thin-wall expansion. This is the same result as what Macpherson and Campbell derived in their paper \cite{Mac}. They concluded further that the neutral F-ball is unstable, taking into account the volume effect proportional to $\epsilon$. We here consider the next-to-leading order contribution in the thin-wall expansion, i.e., the curvature effect, since such effect becomes important when it is comparable to the volume energy effect.

We keep the terms up to the next-to-leading order for $\phi$,
\begin{equation}
\phi=\phi_0+\phi_1~,
\label{neu_phiexpand.eqn}
\end{equation}
with $\phi_1$ smaller than $\phi_0$ by the order of $\delta_n/R$ \footnote{Here $R$ is the order of the curvature radius. Note that the validity of this thin-wall expansion collapses for the small F-ball, $N_f\sim16\pi^{2/3}/\lambda^{2/3}$}. From Eqs.(\ref{neu_eqM.eqn}) and (\ref{neu_bc.eqn}), $\phi_1$ satisfies
\begin{equation}
\ddot{\phi_1}-\frac{\lambda}{2}\left( 3\phi_0^2-v^2 \right)\phi_1=-\left(\frac{1}{R_1}+\frac{1}{R_2}\right)\dot{\phi_0}~,
\label{neu_eqM1.eqn}
\end{equation}
and the boundary condition,
\begin{eqnarray}
\phi_1 = \left\{ 
\begin{array}{ll}
0 & (|w| \gg \delta_n) \\
0 & (w = 0)~~~.
\end{array}
\right.
\label{neu_bc1.eqn}
\end{eqnarray}
The analytic solution \footnote{The solution for $\phi_1$ is not smooth at $w=0$. This is caused by the Yukawa term $-G\bar{\psi}_f\psi_f\phi$ which is implicit in Eq.(\ref{neu_esdef.eqn}). Due to the curvature of the surface, $\bar{\psi}_f\psi_f$ does not vanish and is approximately proportional to $\delta(w)$ which is also implicit in the r.h.s. of Eq.(\ref{neu_eqM1.eqn}), giving the discontinuity of $\dot{\phi}_1$ at the F-ball surface.} for $\phi_1$ is given in Appendix \ref{es1.sec} (see Fig.\ref{phi.fig} for the illustration of $\phi_0$ and $\phi_1$). 
\begin{figure}[htbp]
\includegraphics[height=8cm]{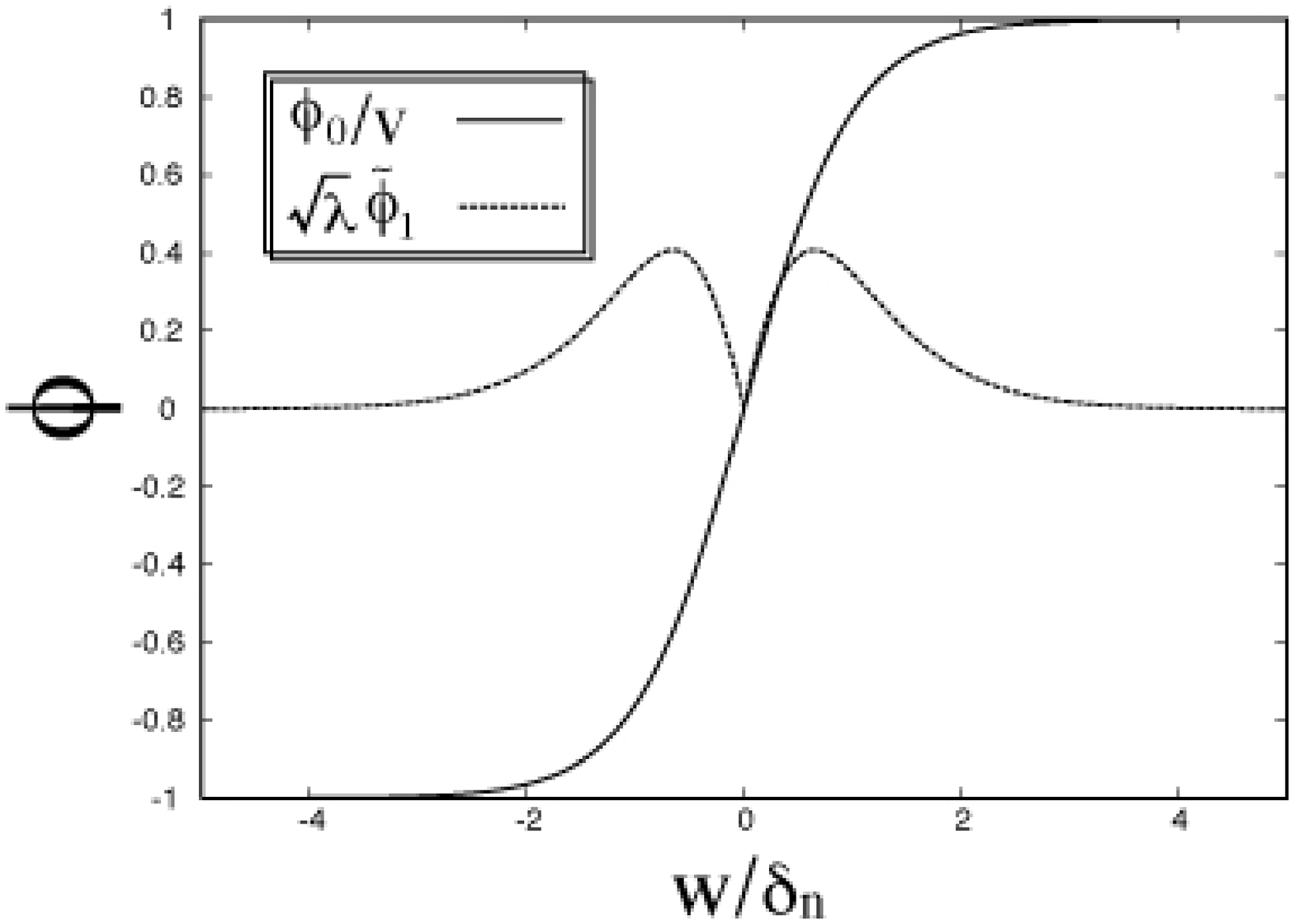}
\caption{The perturbative solutions for the domain wall field $\phi$. Here, $\phi_0$ is the leading term and $\phi_1$ the next-to-leading term in the thin-wall expansion. Note that $\phi_0$ is rescaled as $\phi_0/v$ with $v$ the symmetry breaking scale, and $\phi_1$ as $\sqrt{\lambda}\tilde{\phi_1}$ with $\tilde{\phi_1}=2R_1R_2\phi_1/(R_1+R_2)$ ($R_1$ and $R_2$ are the radii of principal curvature of the F-ball surface).
\label{phi.fig}}
\end{figure}
Substituting the solution for $\phi_1$  and that for $\phi_0$ into $E_{tot}$ and keeping the terms up to the order of $O$($(\delta_n/R)^2E_{tot}^{(0)}$) \footnote{In order to get the order of $(\delta_n/R)^2E_{tot}^{(0)}$ in $E_{tot}$, we need only to keep the order of $(\delta_n/R)\phi_0$ in $\phi$, as is well-known in the perturbation theory.}, we get (see Appendix \ref{es1.sec})
\begin{equation}
E_{tot}=E_{tot}^{(0)}+E_{tot}^{(2)}~,
\label{neu_etot2def.eqn}
\end{equation}
where
\begin{equation}
E_{tot}^{(2)}= C_n\int\mbox{d}S~\frac{1}{\left| R_1R_2 \right|} +D_n\int\mbox{d}S~\left(\frac{1}{R_1}-\frac{1}{R_2}\right)^2\frac{|R_1R_2|}{R_1R_2}~.
\label{neu_etot2.eqn}
\end{equation}
Here, $C_n$ and $D_n$ are given by
\begin{eqnarray}
C_n&=&\int \mbox{d}w~ \left(w^2 \dot{\phi_0}^2-\dot{\phi_0}\tilde{\phi_1} \right) \nonumber \\
&\simeq& -\frac{0.25v}{\sqrt{\lambda}}~, \label{neu_cn.eqn} \\
D_n&=&-\frac{1}{4} \int \mbox{d}w~ \dot{\phi_0}\tilde{\phi_1} \nonumber \\
&\simeq& -\frac{0.28v}{\sqrt{\lambda}}~, \label{neu_dn.eqn}
\end{eqnarray}
with $\tilde{\phi_1}=2R_1R_2\phi_1/(R_1+R_2)$.

First, we examine the stability of the spherical F-ball against the perturbative deformation from the spherical shape (see Fig.\ref{frag.fig}(b)). For this purpose, we take the curvature radius on the F-ball surface positive. In this case, the first term in Eq.(\ref{neu_etot2.eqn}) is constant according to the Gauss-Bonnet theorem \footnote{The integration of the Gaussian curvature on the closed surface takes the topological value, which equals $4\pi$ in this case.}. Thus, the shape-dependent sector in $E_{tot}$ comes from the second term in Eq.(\ref{neu_etot2.eqn}). Because it is semi-negative definite, and zero only for the spherical shape, the spherical F-ball is unstable against the perturbative deformation.

We next consider the stability of the F-ball against fragmentation into smaller ones (see Fig.\ref{frag.fig}(d)). In this case, the fragmentation does not change $E_{tot}^{(0)}$ because it only depends on the fermion number. Here, we compare the energy of the spherical F-ball with that of the fragmented spherical F-balls for simplicity. We see that the second term in Eq.(\ref{neu_etot2.eqn}) vanishes. Thus, only the first term in Eq.(\ref{neu_etot2.eqn}) changes in the fragmentation. Since it is proportional to the number of F-balls due to the Gauss-Bonnet theorem, its negative sign shows that the fragmentation is energetically promoted. Therefore, we find that the F-ball is unstable against not only the deformation from the spherical shape but also the fragmentation into smaller ones.

As pointed out by Macpherson and Campbell, the fragmentation will continue until the thin-wall expansion gets invalid. If the fragmented tiny F-balls sizably contribute to CDM, there should be a large number of them and a large cosmic flux of them as well. The present observations for the dark matter search tell that they can interact with ordinary matter only very weakly. Such F-balls are not considered here since their detectability was discussed in Ref.\cite{Mac}.

\section{STABILITY OF ELECTRICALLY CHARGED F-BALL IN THE UNSCREENED CASE\label{unscreened.sec}}

In the present section, we consider the electrically charged F-ball proposed by Morris \cite{Mor}, in which the fermions trapped on the surface have electric charge.  Assigning the electric charge of $+e$ to the fermion for simplicity, we consider an F-ball carrying the electric charge $+eN_f$.  The energy of the F-ball is expressed as 
\begin{equation}
E_{tot}=E_s+E_f+E_c~,
\label{uns_etot.eqn}
\end{equation}
with $E_c$ the Coulomb potential energy. Since the Coulomb energy is proportional to the fermion number squared $E_c\propto N_f^2$, the total energy per unit fermion number $E_{tot}/N_f$ is an increasing function of $N_f$. Thus, when the F-ball fragments into smaller ones, the total energy apparently decreases. We therefore cannot conclude that the electric force stabilizes the F-ball from fragmentation only by comparing the energies of the two states.  Morris, however, suggested \cite{Mor} that the inclusion of the long ranged repulsive force helps the F-ball to be stabilized. Since it is not so clear whether the electric repulsive force stabilizes the F-ball or not, we here verify it using the perturbative method.

For this purpose, we investigate whether the energy barrier exists between the F-ball before the fragmentation and that after it. Let us consider a spherical F-ball and its perturbative deformation from the spherical shape. Assuming the rotational invariance around $z$-axis for simplicity, we express the surface of the F-ball in the polar coordinates as
\begin{equation}
{\bf x}_f(\theta,\varphi)=r(\theta){\bf e}~,
\label{uns_surf.eqn}
\end{equation}
with ${\bf e}$ a unit vector,
\begin{equation}
{\bf e}=(\sin{\theta}\cos{\varphi},\sin{\theta}\sin{\varphi},\cos{\theta})~.
\label{uns_uvec.eqn}
\end{equation}
We expand $r(\theta)$ as
\begin{eqnarray}
r(\theta)&=&R\left(1+\sum_{l=1}^{\infty}a_lP_l(\cos{\theta})\right) \nonumber\\
&\equiv &R(1+\delta r)~,
\label{uns_rexp.eqn}
\end{eqnarray}
where $a_l$'s are small coefficients of the perturbative expansion, and $ P_l(\cos{\theta}) $ is the Legendre polynomial. We write the surface element as
\begin{equation}
\mbox{d}S=R^2(1+\delta S(a_l)) \mbox{d}\Omega~,
\label{uns_ds.eqn}
\end{equation}
with $\mbox{d}\Omega=\sin{\theta}\mbox{d}\theta\mbox{d}\varphi$. We also expand the surface density of the fermion as
\begin{eqnarray}
\sigma_f(\theta)&=& \frac{N_f}{4 \pi R^2(1+\delta S)} \left(1+ \sum_{l=1}^{\infty} c_l P_l(\cos{\theta})\right) \nonumber \\ 
&\equiv & \frac{N_f}{4 \pi R^2(1+\delta S)} \left(1+\delta \sigma_f \right)~,
\label{uns_nfexp.eqn}
\end{eqnarray}
where $c_l$'s are small expansion coefficients. Note that the above expression satisfies the condition for the fermion number $N_f$ to be conserved (see Eq.(\ref{neu_nftot.eqn})).

Keeping the terms up to the second power of $a_l$ and $c_l$ (see Appendix \ref{unse.sec}), we estimate, within the thin-wall approximation, the surface energy,
\begin{eqnarray}
E_s&=&\Sigma S \nonumber \\
&=&4 \pi \Sigma R^2 \left(1+\sum_{l=1}^{\infty}\frac{h_s^{(l)}}{2l+1} \right)~, \label{uns_es.eqn}
\end{eqnarray}
the fermi energy,
\begin{eqnarray}
E_f&=&\frac{4\sqrt{\pi}}{3}\int\mbox{d}S~\sigma_f^{3/2} \nonumber \\
&=&\frac{2N_f^{3/2}}{3R} \left(1+\sum_{l=1}^{\infty}\frac{h_f^{(l)}}{2l+1} \right)~, \label{uns_ef.eqn}
\end{eqnarray}
and the Coulomb potential energy,
\begin{eqnarray}
E_c&=&\frac{e^2}{8\pi}\int\int\mbox{d}S\mbox{d}S'~ \frac{\sigma_f\sigma_f'}{\left| {\bf x}_f-{\bf x}_f' \right|} \nonumber \\
&=&\frac{e^2N_f^2}{8\pi R} \left(1+\sum_{l=1}^{\infty}\frac{h_c^{(l)}}{2l+1} \right)~, \label{uns_ec.eqn}
\end{eqnarray}
where
\begin{eqnarray}
h_s^{(l)}&=& \left(1+\frac{l(l+1)}{2} \right)a_l^2~, \label{uns_hs.eqn} \\
h_f^{(l)}&=&-\frac{1}{2}\left(1 +\frac{l(l+1)}{2}\right)a_l^2 +\frac{3}{8}b_l^2~, \label{uns_hf.eqn} \\
h_c^{(l)}&=&-\frac{l^2+3l-1}{2l+1}a_l^2 -\frac{2(l-1)}{2l+1}a_lb_l +\frac{1}{2l+1}b_l^2~, \label{uns_hc.eqn}
\end{eqnarray}
with $b_l=c_l-2a_l$. Using the effective radius,
\begin{equation}
R_e \equiv \sqrt{\frac{S}{4\pi}} = R \left(1+\frac{1}{2} \sum_{l=1}^{\infty}\frac{h_s^{(l)}}{2l+1} \right)~,
\label{uns_reff.eqn}
\end{equation}
we express the total energy as
\begin{equation}
E_{tot}=4\pi\Sigma R_e^2 +\frac{2N_f^{3/2}}{3R_e}\left(1+\sum_{l=1}^{\infty}\frac{g_f^{(l)}}{2l+1}\right)+\frac{e^2N_f^2}{8\pi R_e}\left(1+\sum_{l=1}^{\infty}\frac{g_c^{(l)}}{2l+1}\right)~,
\label{uns_e.eqn}
\end{equation}
where
\begin{eqnarray}
g_f^{(l)}&=&\frac{3}{8}b_l^2~, \label{uns_gf.eqn} \\
g_c^{(l)}&=&(l-1)\left(-\frac{3}{2(2l+1)}+\frac{l}{4}\right)a_l^2 -\frac{2(l-1)}{2l+1}a_lb_l +\frac{b_l^2}{2l+1}~. \label{uns_gc.eqn}
\end{eqnarray}
Minimizing $E_{tot}$ with respect to $R_e$, we have
\begin{eqnarray}
E_{tot}&=&(12\pi\Sigma)^{1/3}N_f \left(1+\frac{3e^2\sqrt{N_f}}{16\pi} \right)^{2/3} \nonumber\\
&&\times \left\{1+\frac{2}{3\left(1+\frac{3e^2\sqrt{N_f}}{16\pi} \right)}\sum_{l=1}^{\infty}\frac{1}{2l+1} \left(g_f^{(l)}+\frac{3e^2\sqrt{N_f}}{16\pi}g_c^{(l)} \right) \right\}~,
\label{uns_e_2.eqn}
\end{eqnarray}
with the effective radius,
\begin{eqnarray}
R_e&=&\frac{\sqrt{N_f}}{(12\pi\Sigma)^{1/3}} \left(1+\frac{3e^2\sqrt{N_f}}{16\pi} \right)^{1/3} \nonumber\\
&&\times \left\{1+\frac{1}{3\left(1+\frac{3e^2\sqrt{N_f}}{16\pi} \right)}\sum_{l=1}^{\infty}\frac{1}{2l+1} \left(g_f^{(l)}+\frac{3e^2\sqrt{N_f}}{16\pi}g_c^{(l)} \right) \right\}~.
\label{uns_reff_2.eqn}
\end{eqnarray}

Since the last term in the brackets $\{~\}$ in Eq.(\ref{uns_e_2.eqn}) is semi-positive definite and zero only for $b_1=0$ and $a_l=b_l=0$ for $l\geq 2$, $E_{tot}$ takes local minimum when the F-ball is spherical (we note that the parameter $a_1$ corresponds to the small parallel displacement and not to the deformation of the F-ball). Thus, the spherical F-ball is stable against the perturbative deformation from the spherical shape. It leads to the existence of the energy barrier and should suppress the fragmentation of the spherical F-ball. If the suppression of the fragmentation rate is strong, the F-balls would be effectively stable in the universe (for the detailed discussion on their lifetime, see Appendix \ref{uns_decay.sec})  and could be the candidates for CDM.  However, if $N_f$ is very large, the F-ball should have such a strong electric field that should be quantum-mechanically screened by electrons or positrons with the screening length much shorter than the F-ball radius, as we see in the next section. Therefore, the argument of this section cannot be applied for the F-ball with very large $N_f$.

The argument of this section is applicable only for the F-balls with small $N_f$, which have the radius much smaller than the screening length. Such F-balls should give too large cosmic flux to be compatible with the present observations if the F-balls contribute sizably to CDM (the more details are given in Sec.\ref{constraint.sec} and Ref.\cite{Ara}). We therefore find that the unscreened charged F-ball cannot be a main component of CDM.

\section{STABILITY OF ELECTRICALLY CHARGED F-BALL IN THE SCREENED CASE\label{screened.sec}}

In this section, we examine the stability of such a large F-ball that is electrically charged and screened due to the electrons produced quantum mechanically or thermally.  We call the fermion trapped on the F-ball surface as a 'heavy fermion' in order to distinguish it from the electron; the heavy fermion is almost massless in the two dimensional surface while it is as heavy as $Gv$ in the true vacuum. The property of the screened F-ball is quite different from that of unscreened one, because the long ranged electric force becomes short ranged as a result of the screening.  Let us investigate the screening effect by introducing the Coulomb potential energy $V_e$ for electron ($-V_e$ for positron) with $V_e=-eA_0$. We use the Thomas-Fermi method \cite{Mig,Nas} to deal with this problem \footnote{The Thomas-Fermi approximation is adequate for $|d/dw(1/V_e)| \ll 1$ \cite{Nas}. We see from Eq.(\ref{scr_ve0.eqn}) that this condition is satisfied in the region near the surface where most of the electric charge is screened.}. The Helmholtz free energy is expressed as
\begin{equation}
F_{tot}=F_n+F_{sc}~,
\label{scr_ftotdef.eqn}
\end{equation}
where $F_n$ is the non-Coulombic term that is the same as the energy for the neutral F-ball (see Sec.\ref{neutral.sec}). The second term $F_{sc}$ is the contribution from the screening electrons,
\begin{equation}
F_{sc}=\int\mbox{d}S\mbox{d}w~ \left| (1+\frac{w}{R_1})(1+\frac{w}{R_2}) \right| \left\{ -\frac{1}{2e^2}(\nabla V_e)^2+{\cal F}_e-n_fV_e \right\}~,
\label{scr_fthdef.eqn}
\end{equation}
with ${\cal F}_e$ the free energy density of electron and positron in the Coulomb potential \cite{Kap},
\begin{equation}
{\cal F}_e=-\frac{2T}{(2\pi)^3}\int\mbox{d}^3{\bf p}~ \left\{\log{ \left(1+e^{-\frac{E-V_e}{T}} \right)}+\log{ \left(1+e^{-\frac{E+V_e}{T}} \right)} \right\}~.
\label{scr_fedef.eqn}
\end{equation}
Here, the kinetic energy $E$ is given by $E=\sqrt{{\bf p}^2+m_e^2}$ with $m_e$ the electron mass, and the density of the heavy fermion $n_f$ is given by $n_f=\sigma_f\delta (w)$.  (Note that the negative sign of the first term in $\{~\}$ in Eq.(\ref{scr_fthdef.eqn}) gives the correct electric energy density, $+{\bf E}^2/2$, after the free energy is extremized with respect to $V_e$.)  Taking into account the conservation of the heavy fermion number, we obtain the free energy by extremizing $F_{tot}+\mu(N_f-\int\mbox{d}S~\sigma_f)$ with respect to $\phi$, $\sigma_f$ and $V_e$. The procedure of extremization with respect to $\phi$ and $\sigma_f$ is the same as in Sec.\ref{neutral.sec}, and we refer to the discussions and results there.

Now let us extremize the free energy with respect to $V_e$. Assuming that the variation of $V_e$ in the direction parallel to the F-ball surface is much smaller than that in the normal direction, we get the Thomas-Fermi equation,
\begin{equation}
\frac{1}{e^2} \ddot{V_e}+\frac{1}{e^2} \left( \frac{1}{R_1+w}+\frac{1}{R_2+w} \right)\dot{V_e}=n_f(w)-\overline{n}_e(w)~,
\label{scr_eqM.eqn}
\end{equation}
where the symbol 'dot' stands for the partial derivative with respect to $w$. Here, $\overline{n}_e(w)=\partial {\cal F}_e/\partial V_e$ is the expectation value of the difference between the electron density and the positron density. The second term of l.h.s. in Eq.(\ref{scr_eqM.eqn}) comes from the curvature effect, which was not taken into account in the previous works. We here impose the boundary condition,
\begin{equation}
V_e \rightarrow 0 \hspace{1cm} \mbox{for}~~|w| \gg \delta_{sc}~,
\label{scr_bc.eqn}
\end{equation}
with $\delta_{sc}$ the typical screening length. For the case, $T\gg m_e$, the electron number density is expressed as
\begin{equation}
\overline{n}_e\simeq -\frac{T^2V_e}{6}-\frac{V_e^3}{3\pi^2}~.
\label{scr_nehigh.eqn}
\end{equation}
Note that $\overline{n}_e$ is positive since we take $V_e$ negative, assuming the F-ball carrying the positive electric charge. In the first place, let us obtain the leading contribution in the thin-wall expansion by taking $R_1,R_2\rightarrow \infty$. We get
\begin{equation}
V_e^{(0)}=\frac{-\sqrt{2}\pi T}{\sinh{ \left( \frac{|w|+\delta_{sc}}{\lambda_T} \right)}}~,
\label{scr_ve0.eqn}
\end{equation}
where
\begin{eqnarray}
\lambda_T&=&\frac{\sqrt{3}}{eT}~, \label{scr_lambdat.eqn} \\
\delta_{sc}&=&\lambda_T \mbox{cosh}^{-1}\frac{\sqrt{2}\pi T^2+\sqrt{2\pi^2T^4+3e^2\sigma_f^2}}{\sqrt{3}e\sigma_f} \nonumber \\
&\simeq & \sqrt{\frac{2\sqrt{6}\pi}{e^3\sigma_f}} \hspace{1cm}(v\gg T)~. \label{scr_deltac.eqn}
\end{eqnarray}
Substituting $V_e^{(0)}$ into $F_{sc}$ and keeping the terms of the leading order, we get the leading order of $F_{sc}$ as
\begin{eqnarray}
F_{sc}^{(0)}&=&\int\mbox{d}S\int\mbox{d}w~\left\{-\frac{1}{2e^2}\dot{V}_0^2-\frac{T^2V_0^2}{6}-\frac{V_0^4}{12\pi^2}-\sigma_fV_0\delta(w)\right\} \nonumber \\
&=&\frac{8\pi^2T^3}{\sqrt{3}e}\int\mbox{d}S~\left\{\mbox{coth}^3\left(\frac{\delta_{sc}}{\lambda_T}\right)-\frac{3}{2}\mbox{coth}\left(\frac{\delta_{sc}}{\lambda_T}\right)+\frac{1}{2}\right\} \nonumber \\
&\simeq &\sqrt{\frac{2\sqrt{2}\pi e}{3\sqrt{3}}} \int \mbox{d}S~\sigma_f^{3/2} \hspace{1.5cm}(v\gg T)~.
\label{scr_fth0.eqn}
\end{eqnarray}
From the above equation, we see that the heavy fermions distribute uniformly over the surface, as is the case with Eq.(\ref{neu_uniform_distr.eqn}). Then, $\phi^{(0)}$ in Eq.(\ref{neu_phi0.eqn}) and $V_e^{(0)}$ in Eq.(\ref{scr_ve0.eqn}) extremize the free energy,
\begin{equation}
F_{tot}^{(0)}= \Sigma S+ \left( \frac{4\sqrt{\pi}}{3}+\sqrt{\frac{2\sqrt{2}\pi e}{3\sqrt{3}}} \right) \frac{N_f^{3/2}}{\sqrt{S}}~.
\label{scr_ftot0.eqn}
\end{equation}
Minimizing the free energy with respect to $S$, we obtain
\begin{equation}
F_{tot}^{(0)}=(12\pi\Sigma)^{1/3}\left( 1+\sqrt{\frac{\sqrt{3}e}{4\sqrt{2}}} \right)^{2/3}N_f~,
\label{scr_f0min.eqn}
\end{equation}
with the surface area, 
\begin{equation}
S=\left(\frac{2\sqrt{\pi}}{3\Sigma} \right)^{2/3}\left(1+\sqrt{\frac{\sqrt{3}e}{4\sqrt{2}}} \right)^{2/3}N_f~.
\label{scr_smin.eqn}
\end{equation} 
Thus, we cannot tell, within the leading order of the thin-wall expansion, whether the F-ball is stable against the classical deformation and fragmentation into smaller ones.

\begin{figure}[htbp]
\includegraphics[height=8cm]{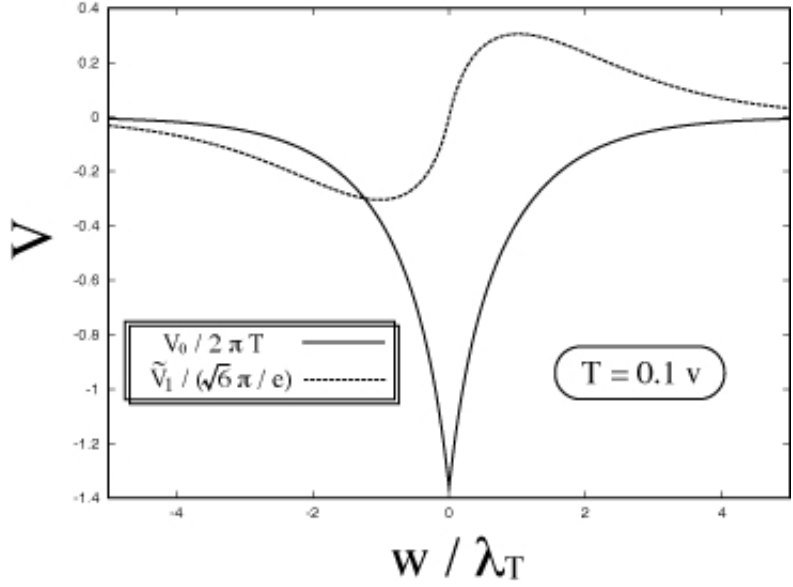}
\caption{The perturbative solutions for the electron potential energy $V_e$. Here, $V_0$ is the leading term and $V_1$ the next-to-leading term in the thin-wall expansion. Note that $V_0$ is rescaled as $V_0/2\pi T$ with $T$ temperature, and $V_1$ as $e\tilde{V_1}/\sqrt{6}\pi$ with $\tilde{V_1}=2R_1R_2V_1/(R_1+R_2)$ ($R_1$ and $R_2$ are the radii of principal curvature of the F-ball surface).
\label{v.fig}}
\end{figure}

We then keep the terms up to the next-to-leading order in the thin-wall expansion to have
\begin{equation}
V_e=V_e^{(0)}+V_e^{(1)}~,
\label{scr_veexpand.eqn}
\end{equation}
with $V_e^{(1)}$ smaller than $V_e^{(0)}$ by the order of $\delta_{sc}/R$ \footnote{Note that the validity of thin-wall expansion collapses for the small F-ball, $N_f\sim 7000$.}. From Eqs.(\ref{scr_eqM.eqn}) and (\ref{scr_bc.eqn}), $V_e^{(1)}$ satisfies
\begin{equation}
\frac{1}{e^2} \ddot{V_e}^{(1)}-\left( \frac{T^2}{3}+\frac{(V_e^{(0)})^2}{\pi^2} \right) V_e^{(1)}=-\frac{1}{e^2} \left( \frac{1}{R_1}+\frac{1}{R_2} \right) \dot{V_e}^{(0)}~,
\label{scr_eqM1.eqn}
\end{equation}
with the boundary condition,
\begin{equation}
V_e^{(1)} \rightarrow 0 \hspace{1cm} \mbox{for}~~|w| \gg \delta_{sc}~.
\label{scr_bc1.eqn}
\end{equation}
The solution to Eq.(\ref{scr_eqM1.eqn}) is given by
\begin{eqnarray}
V_e^{(1)}&=& -\frac{\sqrt{6}\pi|w|}{2ew} \left(\frac{1}{R_1}+\frac{1}{R_2}\right)\frac{\cosh{\left(\frac{|w|+\delta_{sc}}{\lambda_T}\right)}}{\mbox{sinh}^2\left(\frac{|w|+\delta_{sc}}{\lambda_T}\right)} \nonumber \\
&&\hspace{2.5cm}\times \left\{ f\left(\frac{|w|+\delta_{sc}}{\lambda_T}\right)-f\left(\frac{\delta_{sc}}{\lambda_T}\right) \right\}~,
\label{scr_ve1.eqn}
\end{eqnarray}
where
\begin{equation}
f(x)\equiv \frac{1}{3}\sinh{x}\cosh{x}+\frac{2}{3}\tanh{x}-\frac{1}{3}\mbox{sinh}^2x-x
\label{scr_f2.eqn}
\end{equation}
(see Fig.\ref{v.fig} for the illustration of $V_e^{(0)}$ and $V_e^{(1)}$). 
Substituting the solutions, $V_e^{(0)}$ and $V_e^{(1)}$, into $F_{sc}$ and keeping the terms up to the next-to-leading order, we get, in a manner similar to the derivation of Eq.(\ref{neu_etot2def.eqn}),
\begin{equation}
F_{sc}=F_{sc}^{(0)}+F_{sc}^{(2)}~,
\label{scr_fthexpand.eqn}
\end{equation}
where \footnote{In general, we cannot factor $C_{sc}$ and $D_{sc}$ out of the integration because they are dependent of $\sigma_f$. We see, however, that $\sigma_f=const.$ minimizes $F_{tot}^{(0)}$ and that the deviation from the uniform distribution causes the contributions of the order higher than $(\delta_{sc}/R)^2$ in the thin-wall expansion. Thus, we also take $\sigma_f=const.$ in order to estimate $F_{sc}^{(2)}$.}
\begin{equation}
F_{sc}^{(2)}= C_{sc}\int\mbox{d}S~\frac{1}{\left| R_1R_2 \right|} +D_{sc}\int\mbox{d}S~\left(\frac{1}{R_1}-\frac{1}{R_2}\right)^2\frac{|R_1R_2|}{R_1R_2}~.
\label{scr_fth2.eqn}
\end{equation}
Here $C_{sc}$ and $D_{sc}$ are given by
\begin{eqnarray}
C_{sc}&=&-\frac{1}{e^2} \int \mbox{d}w~ \left\{ w^2 (\dot{V_e}^{(0)})^2-\dot{V_e}^{(0)}\tilde{V_e}^{(1)} \right\} \nonumber \\
&\simeq& -50\lambda^{1/6}v~, \label{scr_cth.eqn} \\
D_{sc}&=& \frac{1}{4e^2} \int \mbox{d}w~ \dot{V_e}^{(0)}\tilde{V_e}^{(1)} \nonumber \\
&\simeq& 25\lambda^{1/6}v~, \label{scr_dth.eqn}
\end{eqnarray}
with $\tilde{V_e^{(1)}}=2R_1R_2V_e^{(1)}/(R_1+R_2)$. Note that $D_{sc}$ is positive in contrast to the case with the neutral F-ball where $D_n$ in Eq.(\ref{neu_dn.eqn}) is negative. This feature is very important for the discussion on the stability below. From Eqs.(\ref{neu_etot2.eqn}) and (\ref{scr_fth2.eqn}), we obtain
\begin{eqnarray}
F_{tot}^{(2)}&\simeq &-\left(50\lambda^{1/6}+\frac{0.25}{\sqrt{\lambda}}\right)v\int\mbox{d}S~\frac{1}{\left| R_1R_2 \right|} \nonumber \\
&&+\left(25\lambda^{1/6}-\frac{0.28}{\sqrt{\lambda}}\right)v\int\mbox{d}S~\left(\frac{1}{R_1}-\frac{1}{R_2}\right)^2\frac{|R_1R_2|}{R_1R_2}~.
\label{scr_ftot2.eqn}
\end{eqnarray}

Let us take the curvature radius of the F-ball surface positive, and consider the fragmentation of the spherical F-ball into smaller F-balls. The negative sign of the first term in Eq.(\ref{scr_ftot2.eqn}) shows that the fragmentation is energetically promoted, which is the same as the neutral F-ball in Sec.\ref{neutral.sec}. On the other hand, the sign of the second term is positive for
\begin{equation}
\lambda > 1.2\times 10^{-3}~,
\label{scr_lambda.eqn}
\end{equation}
and there should be the energy barrier to suppress the fragmentation of the F-ball. In this sense, electrically charged F-balls which are screened by electrons are metastable \footnote{We define the F-ball as being metastable when it is classically stable owing to the energy barrier but decays quantum mechanically with a finite lifetime.} at $T\lesssim 0.1v$ (see the discussion in Appendix \ref{s_decay.sec}).

The lifetime of the screened F-ball is estimated in Appendix \ref{s_decay.sec} and is found extremely long, which makes the F-ball effectively stable. Thus, the screened charged F-balls can be the candidates for CDM. In order for them to be a main component of CDM, they should satisfy the constraints to be discussed in the next section.

\section{CONSTRAINTS ON THE PARAMETERS OF F-BALL\label{constraint.sec}}

We consider the phase transition of the universe where the biased $Z_2$ symmetry is spontaneously broken and the network of domain walls is formed \cite{Vil}. The number density of domain walls is soon diluted by annihilation to become $\sim 1/H^3$, where $H$ is the Hubble constant. Using the energy density of the domain walls, $\rho_{wall}\sim\Sigma H$, we get the ratio of $\rho_{wall}$ to the total energy density in the universe $\rho_{tot}$,
\begin{equation}
\frac{\rho_{wall}}{\rho_{tot}} \simeq \frac{\Sigma H}{g_*T^4} \simeq \frac{\Sigma}{H M_{pl}^2}~,
\label{con_rhowall_rhotot.eqn}
\end{equation}
with $g_*$ the effective degree of freedom \cite{Kol} and $M_{pl}$ the Planck mass. The ratio will be of the order of unity when $H\sim H_{eq}=\Sigma/M_{pl}^2$.

The small violation of the $Z_2$ symmetry gives an energy of the false vacuum bubble proportional to its volume. This volume energy increases with the universe expansion and gets comparable to the surface energy at $H\sim H_f=\epsilon/\Sigma$. The domain walls then stop to expand getting round to produce F-balls. From the cosmological viewpoint, F-balls should be produced before the domain walls dominate the total energy density in the universe, since otherwise the energy of the universe would be dominated by blackholes made with the domain walls \cite{Vil}. This condition is $H_f>H_{eq}$, which is rewritten in terms of $\epsilon$,
\begin{equation}
\epsilon > 0.4\lambda v^6 M_{pl}^{-2}~.
\label{con_epsilon_l.eqn}
\end{equation}

Let us consider the case where the F-balls are metastable and may contribute sizably to CDM in the present universe. As mentioned in Sec.\ref{neutral.sec}, we are not interested in the neutral F-ball, which is very tiny after the fragmentations,  since we need too many new assumptions in introducing such an object and have only indirect means to detect it.  We thus consider the charged F-balls. Since they interact with ordinary matter through the electric force, they would be easily observed in various experiments for the dark matter search if there exists enough cosmic flux. If the F-ball has a considerable contribution to CDM, the number density and its cosmic flux should be inversely proportional to its mass, $1/M_f$.  Since such events have not been observed so far, we should have $M_f>10^{25}$ GeV \cite{Ara}. This constraint is rewritten in terms of $N_f$,
\begin{equation}
N_f > 10^{25}\left( \frac{\mbox{GeV}}{\lambda^{1/6} v} \right)~.
\label{con_nf.eqn}
\end{equation}
If the size of the F-ball is smaller than the typical screening length, its electric charge is not screened at the close neighborhood of the F-ball but screened far outside of it. In this case, the discussion on the stability made in Sec.\ref{unscreened.sec} is valid. From Eqs.(\ref{uns_reff_2.eqn}) and (\ref{scr_deltac.eqn}), we have the size of the unscreened F-ball $R\sim 0.5\sqrt{N_f}(1+4/3\alpha \sqrt{N_f})^{1/3}/\lambda^{1/6}v$ and the screening length $\delta_{sc}\sim 30/\lambda^{1/6}v$, which gives a very small number of the fermion on the unscreened F-ball, $N_f<1000$, from $R<\delta_{sc}$. This is incompatible with Eq.(\ref{con_nf.eqn}), and we see that the unscreened F-ball cannot give a sizable contribution to CDM.

If the size of the F-ball is much larger than the screening length (i.e., $R\gg \delta_{sc}$), its electric charge is strongly screened as was discussed in the previous section. In this case, Eqs.(\ref{scr_deltac.eqn}) and (\ref{scr_smin.eqn}) give $N_f\gg 7000$, which is compatible with Eq.(\ref{con_nf.eqn}). This allows screened charged F-balls to be candidates for the main component of CDM in the present universe.

In the previous section, we found that the second term in Eq.(\ref{scr_ftot2.eqn}) is crucial to enhancing the stability and suppressing the fragmentation of the screened charged F-ball for the parameter satisfying Eq.(\ref{scr_lambda.eqn}). If we add the volume energy $E_v\sim \epsilon V$ which has the destabilizing effect \footnote{It is easy to understand within the leading order discussion in the thin-wall expansion \cite{Mac,Mor}.}, we still have the metastable F-ball as far as the second term in Eq.(\ref{scr_ftot2.eqn}) overwhelms $E_v$. Using Eq.(\ref{scr_dth.eqn}), we express this condition as
\begin{equation}
\epsilon < 100\lambda^{2/3}v^4N_f^{-3/2}~.
\label{con_epsilon_u.eqn}
\end{equation}
The equations (\ref{con_epsilon_l.eqn}) to (\ref{con_epsilon_u.eqn}) and Eq.(\ref{scr_lambda.eqn}) give the following condition for the symmetry breaking scale,
\begin{equation}
v < 3\times 10^{6}~\mbox{GeV}~.
\label{con_v.eqn}
\end{equation}

\section{SUMMARY AND DISCUSSION\label{conclusion.sec}}

We have considered the F-balls produced in the early universe due to the spontaneous breakdown of the biased $ Z_2 $ symmetry. We have investigated the stability of the electrically neutral F-balls and the charged F-balls in order to examine whether they can sizably contribute to CDM or not.

In the case of neutral F-balls, we have taken into account the correction to the thin-wall approximation up to the next-to-leading order, and found this correction plays an important role to enhance deformation and fragmentation of the F-ball into tiny thick-wall F-balls. If these tiny F-balls are a main component of CDM, the dark matter search experiments allow the F-ball to have only a weak interaction with the ordinary matter. We are not interested in such a new kind of weakly interacting particles since we need too many new assumptions in introducing new tiny particles and have only indirect means to detect such particles \cite{Mac}.

In the case of charged F-balls, two cases are considered:

case 1) the size of a charged F-ball being smaller than the typical screening length. Such F-balls are found to have rather light masses,  and they have a large number density in the universe if they are a main component of CDM.  However, since the cosmic flux and the number density of the charged F-balls are observed to be very small, it is difficult for them to be a main component of CDM.  (In this case, the electric charge of the F-balls is effectively unscreened. We have found that they decay into smaller F-balls through the tunneling effect  though they are classically stable. Since this decay further increases the large number density of the F-balls, it merely decreases the possibility for them to be a main component of CDM.)

case 2) the size of a charged F-ball being larger than the typical screening length. In this case, the electric charge of the F-ball is screened in the vicinity of its surface. Even though the long-range character of the Coulomb force is lost, the F-balls can still be classically stable in certain region of the parameters. This is due to the curvature effect that is obtained from the next-to-leading order correction to the thin-wall approximation. Such F-balls can be candidates for a main part of CDM in the present universe. In such a case, we have obtained constraints on the physical parameters: $\lambda > 1.2\times 10^{-3}$, $0.4\lambda v^6M_{pl}^{-2}<\epsilon <100\lambda^{2/3}v^4N_f^{-3/2}$ and $v<3\times 10^6$ GeV. It is interesting to note that the symmetry breaking scale $v$ constrained above is not too far from Electroweak or SUSY breaking scale. These above constraints help us to make a realistic model of the F-ball.

Recently the dark matter distributions of galaxies and clusters were observed and compared with the dark matter simulations.  Some find consistency of the collisionless CDM simulation with the observations \cite{Kra, Ogu}, while others find inconsistency and propose \cite{Spe} the self-interacting dark matter (SIDM) with the ratio of the cross section to the mass as large as $10^{-23}-10^{-24}\mbox{cm}^2\mbox{GeV}^{-1}$. In the case of screened charged F-ball, the geometrical cross section per unit mass is not so large,
\begin{equation}
\frac{\sigma_{g}}{M_f}\simeq 0.5\lambda^{-1/2}\times 10^{-28}~\left(\frac{\mbox{GeV}}{v}\right)^3~\mbox{cm}^2\mbox{GeV}^{-1}~,
\label{con_sigmaM.eqn}
\end{equation}
and, moreover, the typical momentum transfer squared $\sim 1/R^2 \sim v^2/N_f$ is too small to give a large angle scattering which appears to be necessary for smoothing the Halo central density profile. Thus, the present model of the F-ball is not adequate for SIDM. We, however, do not discuss this issue further in the present paper since the above problem is still controversial \cite{Kra, Tys, Ste}.

All through the present paper, we have dealt with the heavy fermions as being distributed two-dimensionally on the F-ball surface, and neglected the spreading of the fermion in the direction normal to the surface. This spreading may have some contributions to the curvature effect energy of the F-ball. We will discuss this issue elsewhere.

\appendix

\section{VOLUME ELEMENT IN NEW COORDINATES\label{dv.sec}}

We introduce three local orthogonal coordinates ($u,v,w$) for the F-ball, the origin of which is on the surface of an F-ball, and derive a representation for the volume element in this frame. The coordinates, $u$ and $v$, represent the F-ball surface, ${\bf x}={\bf x}_f(u,v)$, and are taken along the lines of curvature \footnote{Two lines, $u=\mbox{const.}$ and $v=\mbox{const.}$, are called 'lines of curvature' if their tangent vectors are parallel to the principal directions.} on the surface. We define the first and the second fundamental forms,
\begin{eqnarray}
I&=&E(u,v) \mbox{d}u^2+2F(u,v) \mbox{d}u \mbox{d}v+G(u,v) \mbox{d}v^2~,
\label{dv_def1.eqn} \\
J&=&L(u,v) \mbox{d}u^2+2M(u,v) \mbox{d}u \mbox{d}v+N(u,v) \mbox{d}v^2~,
\label{dv_def2.eqn}
\end{eqnarray}
with the coefficients,
\begin{eqnarray}
E(u,v)&=&{\bf x}_{f,u}{\bf x}_{f,u}~, \hspace{1cm} L(u,v)={\bf x}_{f,uu}{\bf n}~, \nonumber \\
F(u,v)&=&{\bf x}_{f,u}{\bf x}_{f,v}~, \hspace{1cm} M(u,v)={\bf x}_{f,uv}{\bf n}~, \nonumber \\
G(u,v)&=&{\bf x}_{f,v}{\bf x}_{f,v}~, \hspace{1cm} N(u,v)={\bf x}_{f,vv}{\bf n}~,
\label{dv_coe.eqn}
\end{eqnarray}
where the subscripts of $u$ and $v$ stand for the partial derivative with respect to them, and $ {\bf n}(u,v) $ is a unit vector normal to the surface defined as
\begin{equation}
{\bf n}(u,v)=\frac{{\bf x}_{f,u} \times {\bf x}_{f,v}}{\left| {\bf x}_{f,u} \times {\bf x}_{f,v} \right|}~.
\label{dv_normal_vec.eqn}
\end{equation}
Since we take $u=$const. and $v=$const. as the lines of curvature, we obtain
\begin{equation}
F(u,v)=M(u,v)=0~.
\label{dv_curv_line.eqn}
\end{equation}
We define the third coordinate $w$ \footnote{The coordinates can be defined uniquely as far as two or more normal lines do not intersect. We can neglect the contribution in the energy integration over such a region of intersection, since the region is far from the surface.} by
\begin{equation}
{\bf x}(u,v,w)={\bf x}_f(u,v)+w{\bf n}(u,v)~,
\label{dv_defcoord.eqn}
\end{equation}
taking $w$ positive (negative) for the outside (inside) of the F-ball. We take $w=0$ at the place in the wall where the scalar field $\phi$ vanishes.

Using the orthogonal relations,
\begin{equation}
{\bf x}_{f,u}{\bf x}_{f,v}={\bf x}_{f,u}{\bf n}={\bf x}_{f,v}{\bf n}={\bf n}\mbox{d}{\bf n}=0~,
\label{dv_orthogonal.eqn}
\end{equation}
we express the line element squared,
\begin{eqnarray}
\mbox{d}^2{\bf x}&=&({\bf x}_{f,u} \mbox{d}u+{\bf x}_{f,v} \mbox{d}v+{\bf n} \mbox{d}w+w \mbox{d}{\bf n})^2 \nonumber \\
&=&E\mbox{d}u^2+G\mbox{d}v^2+2w({\bf x}_{f,u}\mbox{d}{\bf n} \mbox{d}u+{\bf x}_{f,v}\mbox{d}{\bf n} \mbox{d}v)+w^2\mbox{d}{\bf n}^2+\mbox{d}w^2~.
\label{dv_dx2.eqn}
\end{eqnarray}
From Eq.(\ref{dv_orthogonal.eqn}) and $ M(u,v)=0 $ in Eq.(\ref{dv_curv_line.eqn}), we obtain
\begin{eqnarray}
{\bf n}_u &=& -\frac{L}{E}{\bf x}_{f,u}~, \label{dv_nu.eqn} \\
{\bf n}_v &=& -\frac{N}{G}{\bf x}_{f,v}~. \label{dv_nv.eqn}
\end{eqnarray}
Thus, we obtain
\begin{eqnarray}
{\bf x}_{f,u}\mbox{d}{\bf n} &=& {\bf x}_{f,u} \left( {\bf n}_u \mbox{d}u+{\bf n}_v \mbox{d}v \right) = -L \mbox{d}u~, \label{dv_xudn.eqn} \\
{\bf x}_{f,v}\mbox{d}{\bf n} &=& {\bf x}_{f,v} \left( {\bf n}_u \mbox{d}u+{\bf n}_v \mbox{d}v \right) = -N \mbox{d}v~, \label{dv_xvdn.eqn} \\
\mbox{d}{\bf n}^2 &=& \left( {\bf n}_u \mbox{d}u+{\bf n}_v \mbox{d}v \right)^2 \nonumber \\
&=& \frac{L^2}{E}\mbox{d}u^2+\frac{N^2}{G}\mbox{d}v^2~. \label{dv_dn2.eqn}
\end{eqnarray}
Substituting Eqs.(\ref{dv_xudn.eqn}) to (\ref{dv_dn2.eqn}) into Eq.(\ref{dv_dx2.eqn}), we get
\begin{equation}
\mbox{d}^2{\bf x}=E \left( 1-\frac{wL}{E} \right)^2 \mbox{d}u^2+G \left( 1-\frac{wN}{G} \right)^2 \mbox{d}v^2+ \mbox{d}w^2~.
\label{dv_dx2_2.eqn}
\end{equation}

Calculating a determinant of the line element squared, we obtain the volume element,
\begin{eqnarray}
\mbox{d}^3{\bf x} &=& \sqrt{EG \left( 1-\frac{wL}{E} \right)^2 \left( 1-\frac{wN}{G} \right)^2} \mbox{d}u \mbox{d}v \mbox{d}w \nonumber \\
&=& \left| \left( 1-\frac{wL}{E} \right) \left( 1-\frac{wN}{G} \right) \right| \mbox{d}S \mbox{d}w~,
\label{dv_dx3.eqn}
\end{eqnarray}
where we use the relation,
\begin{equation}
\mbox{d}S = \sqrt{EG} \mbox{d}u \mbox{d}v~.
\label{dv_ds.eqn}
\end{equation}
Taking $R_1$ and $R_2$ as the radii of principal curvature of the surface, we obtain the Gaussian curvature,
\begin{equation}
K \equiv \frac{1}{R_1R_2}=\frac{LN}{EG}~,
\label{dv_gaussian.eqn}
\end{equation}
the mean curvature,
\begin{equation}
H \equiv \frac{1}{2} \left( \frac{1}{R_1}+\frac{1}{R_2} \right)=-\frac{EN+GL}{2EG}~,
\label{dv_mean.eqn}
\end{equation}
and the volume element,
\begin{equation}
\mbox{d}^3{\bf x}=\left| (1+\frac{w}{R_1})(1+\frac{w}{R_2}) \right| \mbox{d}S \mbox{d}w~.
\label{dv_dv.eqn}
\end{equation}

\section{CURVATURE EFFECT IN NEUTRAL CASE\label{es1.sec}}

We here derive Eq.(\ref{neu_etot2def.eqn}) to (\ref{neu_dn.eqn}) in Sec.\ref{neutral.sec}, assuming $R_1,~R_2>0$. Noting that higher order effects in $E_{tot}$ arise only from $E_s$, we estimate $E_s$ keeping the terms up to the order of $(\delta_n/R)^2$. Substituting Eq.(\ref{neu_phiexpand.eqn}) into Eq.(\ref{neu_esdef.eqn}), we obtain
\begin{equation}
E_s=E_s^{(0)}+E_s^{(1)}+E_s^{(2)}~,
\label{es_esdef.eqn}
\end{equation}
where
\begin{eqnarray}
E_s^{(0)}&=&\int\mbox{d}S\int_{-\infty}^{+\infty}\mbox{d}w~\left\{\frac{1}{2}\dot{\phi_0}^2+\frac{\lambda}{8}(\phi_0^2-v^2)^2\right\}~, \label{es_es0def.eqn} \\
E_s^{(1)}&=&\int\mbox{d}S\int_{-\infty}^{+\infty}\mbox{d}w~\left[\left(\frac{w}{R_1}+\frac{w}{R_2}\right)\left\{\frac{1}{2}\dot{\phi_0}^2+\frac{\lambda}{8}(\phi_0^2-v^2)^2\right\}\right. \nonumber \\
&&\hspace{2cm}+\left.\left\{\dot{\phi_0}\dot{\phi_1}+\frac{\lambda}{2}\phi_0(\phi_0^2-v^2)\phi_1\right\}\right]~, \label{es_es1def.eqn} \\
E_s^{(2)}&=&\int\mbox{d}S\int_{-\infty}^{+\infty}\mbox{d}w~\left[\frac{w^2}{R_1R_2}\left\{\frac{1}{2}\dot{\phi_0}^2+\frac{\lambda}{8}(\phi_0^2-v^2)^2\right\}\right. \nonumber \\
&&\hspace{2cm}+\left(\frac{w}{R_1}+\frac{w}{R_2}\right)\left\{\dot{\phi_0}\dot{\phi_1}+\frac{\lambda}{2}\phi_0(\phi_0^2-v^2)\phi_1\right\} \nonumber \\
&&\hspace{4cm}+\left.\left\{\frac{1}{2}\dot{\phi_1}^2+\frac{\lambda}{4}(3\phi_0^2-v^2)\phi_1^2\right\}\right]~. \label{es_es2def.eqn}
\end{eqnarray}
Substituting Eq.(\ref{neu_phi0.eqn}) into Eq.(\ref{es_es0def.eqn}), we get
\begin{equation}
E_s^{(0)}=\Sigma \int\mbox{d}S~,
\label{es_es0.eqn}
\end{equation}
with $\Sigma=2\sqrt{\lambda}v^3/3$. The integration of Eq.(\ref{es_es1def.eqn}) vanishes,
\begin{equation}
E_s^{(1)}=0~,
\label{es_es1.eqn}
\end{equation}
because the integrand is an odd function of $w$. Using the relation, $\dot{\phi_0}^2=\lambda(\phi_0^2-v^2)^2/4$ derived from Eq.(\ref{neu_eqM0.eqn}), we express the first line in the r.h.s. of Eq.(\ref{es_es2def.eqn}) as
\begin{equation}
(\mbox{1st~line})=\int\mbox{d}S\int_{-\infty}^{+\infty}\mbox{d}w~\frac{w^2}{R_1R_2}\dot{\phi_0}^2~.
\label{es_es2first.eqn}
\end{equation}
The second line is rewritten as
\begin{eqnarray}
(\mbox{2nd~line})&=&\int\mbox{d}S\int_{-\infty}^{+\infty}\mbox{d}w~\left[\left(\frac{w}{R_1}+\frac{w}{R_2}\right)\left\{-\ddot{\phi_0}+\frac{\lambda}{2}\phi_0(\phi_0^2-v^2)\right\}\phi_1\right. \nonumber \\
&&\hspace{2cm}-\left.\left(\frac{1}{R_1}+\frac{1}{R_2}\right)\dot{\phi_0}\phi_1\right] \nonumber \\
&=&-\int\mbox{d}S\int_{-\infty}^{+\infty}\mbox{d}w~\left(\frac{1}{R_1}+\frac{1}{R_2}\right)\dot{\phi_0}\phi_1~. \label{es_es2second.eqn}
\end{eqnarray}
From Eq.(\ref{neu_eqM1.eqn}), we get for the third line,
\begin{eqnarray}
(\mbox{3rd~line})&=&\int\mbox{d}S\int_{-\infty}^{+\infty}\mbox{d}w~\left\{-\frac{1}{2}\ddot{\phi_1}+\frac{\lambda}{4}(3\phi_0^2-v^2)\phi_1\right\}\phi_1 \nonumber \\
&=&\frac{1}{2}\int\mbox{d}S\int_{-\infty}^{+\infty}\mbox{d}w~\left(\frac{1}{R_1}+\frac{1}{R_2}\right)\dot{\phi_0}\phi_1~. \label{es_es2third.eqn}
\end{eqnarray}
From Eq.(\ref{es_es2first.eqn}) to (\ref{es_es2third.eqn}), we obtain
\begin{equation}
E_s^{(2)}=\int\mbox{d}S\int_{-\infty}^{+\infty}\mbox{d}w~\left\{\frac{w^2}{R_1R_2}\dot{\phi_0}^2-\frac{1}{2}\left(\frac{1}{R_1}+\frac{1}{R_2}\right)\dot{\phi_0}\phi_1\right\}~.
\label{es_es2.eqn}
\end{equation}
Using the dimensionless quantity,
\begin{equation}
\tilde{\phi_1}\equiv \frac{2R_1R_2}{R_1+R_2}\phi_1~,
\label{es_phi1tildedef.eqn}
\end{equation}
we express $E_s^{(2)}$ as
\begin{eqnarray}
E_s^{(2)}&=&\int\mbox{d}S\int_{-\infty}^{+\infty}\mbox{d}w~\left\{\frac{w^2}{R_1R_2}\dot{\phi_0}^2-\frac{1}{4}\left(\frac{1}{R_1}+\frac{1}{R_2}\right)^2\dot{\phi_0}\tilde{\phi_1}\right\} \nonumber \\
&=&C_n\int\mbox{d}S~\frac{1}{R_1R_2} +D_n\int\mbox{d}S~\left(\frac{1}{R_1}-\frac{1}{R_2}\right)^2~, \label{es_es2final.eqn}
\end{eqnarray}
with
\begin{eqnarray}
C_n&=&\int_{-\infty}^{+\infty}\mbox{d}w~ \left(w^2 \dot{\phi_0}^2-\dot{\phi_0}\tilde{\phi_1} \right)~, \label{es_cndef.eqn} \\
D_n&=&-\frac{1}{4} \int_{-\infty}^{+\infty}\mbox{d}w~ \dot{\phi_0}\tilde{\phi_1}~. \label{es_dndef.eqn}
\end{eqnarray}
From Eq.(\ref{neu_eqM1.eqn}) with Eq.(\ref{neu_bc1.eqn}), we get
\begin{eqnarray}
\tilde{\phi_1}(w)&=&\frac{1}{\sqrt{\lambda}}\mbox{sech}^2\left(\frac{w}{\delta_n}\right)-\frac{1}{\sqrt{\lambda}}\mbox{cosh}^2\left(\frac{w}{\delta_n}\right)+\frac{1}{3\sqrt{\lambda}}\mbox{sinh}^2\left(\frac{w}{\delta_n}\right) \mbox{tanh}^2\left(\frac{w}{\delta_n}\right) \nonumber \\
&&+\frac{1}{12\sqrt{\lambda}}\mbox{sech}^2\left(\frac{w}{\delta_n}\right) \left| \sinh{\left(\frac{4w}{\delta_n}\right)}+8 \sinh{\left(\frac{2w}{\delta_n}\right)}+\frac{12w}{\delta_n}\right|~,
\label{es_phi1tilde.eqn}
\end{eqnarray}
with $\delta_n=2/\sqrt{\lambda}v$. Substituting Eqs.(\ref{neu_phi0.eqn}) and (\ref{es_phi1tilde.eqn}) into Eqs.(\ref{es_cndef.eqn}) and (\ref{es_dndef.eqn}), we obtain
\begin{eqnarray}
C_n&=&\frac{v}{\sqrt{\lambda}}\int_{0}^{\infty}\mbox{d}\tilde{w}~\left\{4\tilde{w}^2\mbox{sech}^4\tilde{w}-2\mbox{sech}^4\tilde{w}+2-\frac{2}{3}\mbox{tanh}^4\tilde{w}\right. \nonumber \\
&&\hspace{2cm}-\left.\frac{1}{6}\mbox{sech}^4\tilde{w}(\mbox{sinh}4\tilde{w}+8\mbox{sinh}2\tilde{w}+12\tilde{w})\right\} \nonumber \\
&=&\frac{v}{\sqrt{\lambda}}\left(\int_{0}^{\infty}\mbox{d}\tilde{w}~4\tilde{w}^2\mbox{sech}^4\tilde{w}-\frac{10}{9}\right)\simeq -\frac{0.25v}{\sqrt{\lambda}}~, \label{es_cn.eqn} \\
D_n&=&\frac{v}{\sqrt{\lambda}}\int_{0}^{\infty}\mbox{d}\tilde{w}~\left\{-\frac{1}{2}\mbox{sech}^4\tilde{w}+\frac{1}{2}-\frac{1}{6}\mbox{tanh}^4\tilde{w}\right. \nonumber \\
&&\hspace{2cm}-\left.\frac{1}{24}\mbox{sech}^4\tilde{w}(\mbox{sinh}4\tilde{w}+8\mbox{sinh}2\tilde{w}+12\tilde{w})\right\} \nonumber \\
&=&-\frac{5v}{18\sqrt{\lambda}}~. \label{es_dn.eqn}
\end{eqnarray}

\section{ENERGY OF UNSCREENED F-BALL\label{unse.sec}}

We derive the thin-wall representations for the energy of an unscreened charged F-ball, Eqs.(\ref{uns_es.eqn}), (\ref{uns_ef.eqn}) and (\ref{uns_ec.eqn}). We first consider the surface energy,
\begin{equation}
E_s=\Sigma \int\mbox{d}S~.
\label{unse_esdef.eqn}
\end{equation}
Using Eq.(\ref{uns_surf.eqn}), we get the metric of the surface,
\begin{equation}
\mbox{d}{\bf x}^2=\left(r^2+\dot{r}^2\right)\mbox{d}\theta^2+r^2\mbox{sin}^2\theta\mbox{d}\varphi^2~,
\label{unse_metric.eqn}
\end{equation}
with the symbol 'dot' standing for the derivative with respect to $\theta$. From Eq.(\ref{unse_metric.eqn}), we obtain
\begin{equation}
\mbox{d}S=r^2\sqrt{1+\left(\frac{\dot{r}}{r}\right)^2}\mbox{d}\Omega~.
\label{unse_ds.eqn}
\end{equation}
Substituting $\mbox{d}S$ in Eq.(\ref{unse_ds.eqn}) into Eq.(\ref{unse_esdef.eqn}), with $r$ expressed in Eq.(\ref{uns_rexp.eqn}), and using the following orthogonality relations,
\begin{eqnarray}
\int\frac{\mbox{d}\Omega}{4\pi}P_lP_{m}&=&\frac{1}{2l+1}\delta_{lm}~, \label{unse_formula01.eqn} \\
\int\frac{\mbox{d}\Omega}{4\pi}\dot{P}_l\dot{P}_{m}&=&\frac{l(l+1)}{2l+1}\delta_{lm}~,
\label{unse_formula02.eqn}
\end{eqnarray}
we obtain Eq.(\ref{uns_es.eqn}) with Eq.(\ref{uns_hs.eqn}).

We next consider the fermi energy,
\begin{equation}
E_f=\frac{4\sqrt{\pi}}{3}\int\mbox{d}S~\sigma_f^{3/2}~.
\label{unse_efdef.eqn}
\end{equation}
Using Eq.(\ref{uns_nfexp.eqn}) and Eq.(\ref{unse_ds.eqn}), we rewrite Eq.(\ref{unse_efdef.eqn}) as
\begin{eqnarray}
E_f&=&\frac{2N_f^{3/2}}{3R}\int\frac{\mbox{d}\Omega}{4\pi}\frac{(1+\delta \sigma_f)^{3/2}}{(1+\delta S)^{1/2}} \nonumber\\
&=&\frac{2N_F^{3/2}}{3R}\int\frac{\mbox{d}\Omega}{4\pi}(1+\delta F)~,
\label{unse_ef01.eqn}
\end{eqnarray}
where
\begin{equation}
\delta F=\frac{3}{2}\delta \sigma_f+\frac{3}{8}\delta \sigma_f^2-\delta r+\delta r^2-\frac{1}{4}\delta {\dot r}^2-\delta \sigma_f \delta r~.
\label{unse_df.eqn}
\end{equation}
This gives Eq.(\ref{uns_ef.eqn}) with Eq.(\ref{uns_hf.eqn}).

We finally consider the Coulomb energy,
\begin{equation}
E_c=\frac{e^2}{8\pi}\int\int\mbox{d}S\mbox{d}S'~ \frac{\sigma_f\sigma_f'}{\left| {\bf x}-{\bf x}' \right|}~.
\label{unse_ecdef.eqn}
\end{equation}
Using the expression,
\begin{equation}
\frac{1}{4\pi|{\bf x}-{\bf x}'|}=\int\frac{\mbox{d}^3{\bf k}}{(2\pi)^3k^2}e^{i{\bf k}({\bf x}-{\bf x}')}~,
\label{unse_formula03.eqn}
\end{equation}
we rewrite Eq.(\ref{unse_ecdef.eqn}) as
\begin{equation}
E_c=\frac{e^2N_F^2}{4\pi^2}\int_{0}^{\infty}\mbox{d}k\int\frac{\mbox{d}\Omega_k}{4\pi}\left| \int\frac{\mbox{d}\Omega_{x}}{4\pi}~\left(1+\delta \sigma_f \right)e^{i{\bf k}{\bf x}}\right|^2~.
\label{unse_ec01.eqn}
\end{equation}
Using the following expansion relation,
\begin{eqnarray}
e^{i{\bf k}{\bf x}}&=&\sum_{l=0}^{\infty}i^l j_l(kr) (2l+1)P_l(\cos \Theta_{kx}) \nonumber\\
&=&\sum_{l=0}^{\infty}\sum_{m} 4\pi i^l j_l(kr) Y_{lm}(\Omega_k)Y_{lm}^\ast(\Omega_x)
\label{unse_formula04.eqn}
\end{eqnarray}
(here $j_l(kr)$ and $Y_{lm}(\Omega)$ are the spherical Bessel function and the spherical harmonics, respectively, and $\Theta_{kx}$ is the angle between {\bf k} and {\bf x}), and the orthogonality relation,
\begin{equation}
\int\mbox{d}\Omega_k Y_{lm}(\Omega_k)Y_{l'm'}^\ast(\Omega_k)=\delta_{ll'}\delta_{mm'}~,
\label{unse_formula05.eqn}
\end{equation}
we obtain
\begin{equation}
E_c=\frac{e^2N_F^2}{4\pi^2}\sum_{l=0}^{\infty}\frac{D_c^{(l)}}{2l+1}~,
\label{unse_ec02.eqn}
\end{equation}
where
\begin{eqnarray}
D_c^{(l)}&=& (2l+1)^2\lim_{\epsilon \rightarrow 0} \int_0^\infty\mbox{d}ke^{-\epsilon k}\int\frac{\mbox{d}(\cos{\theta})}{2}\sigma_f\int\frac{\mbox{d}(\cos{\theta'})}{2}\sigma_f' \nonumber\\
&& \hspace{2.5cm}\times j_l(kr)j_l(kr')P_l(\cos{\theta})P_l(\cos{\theta'})~.
\label{unse_dcdef.eqn}
\end{eqnarray}
Here, we have introduced the procedure, $ \lim_{\epsilon \rightarrow 0} $, in order to safely expand the spherical Bessel function in the integrand,
\begin{equation}
j_l(kr) \simeq j_l(kR) + \delta r kR j_l'(kR) +\frac{1}{2} \delta r^2 (kR )^2 j_l''(kR)~.
\label{unse_besselexp.eqn}
\end{equation}
Keeping the terms up to the second order, we obtain
\begin{eqnarray}
D_c^{(l)}&=&(2l+1)^2\lim_{\epsilon \rightarrow 0} \int_0^\infty\mbox{d}ke^{-\epsilon k}\int\frac{\mbox{d}(\cos{\theta})}{2}P_l(\cos{\theta}) \nonumber\\
&& \hspace{2cm}\times\int\frac{\mbox{d}(\cos{\theta'})}{2}P_l(\cos{\theta'}) I_1^{(l)}(\theta,\theta')~,
\label{unse_dc01.eqn}
\end{eqnarray}
where
\begin{eqnarray}
I_1^{(l)}&=& (1+\delta \sigma_f + \delta \sigma_f' +\delta \sigma_f \delta \sigma_f')j_l(kR)^2 \nonumber\\
&&+(1+ \delta \sigma_f + \delta \sigma_f')(\delta r + \delta r') kR j_l(kR) j_l'(kR) \nonumber\\
&&+ \delta r \delta r'(kR)^2 (j_l'(kR))^2 +\delta r \delta r' (kR)^2 (j_l'(kR))^2 \nonumber\\
&&+\frac{1}{2}(\delta r^2 + \delta r'^2 )(kR)^2 j_l(kR)j_l''(kR)~.
\label{unse_i1.eqn}
\end{eqnarray}
Integrating Eq.(\ref{unse_dc01.eqn}) with respect to the angular variables, we get
\begin{equation}
D_c^{(l)}= \lim_{\epsilon \rightarrow 0} \int_0^\infty \mbox{d}k e^{-\epsilon k}  I_2^{(l)}~,
\label{unse_dc02.eqn}
\end{equation}
where
\begin{eqnarray}
I_2^{(l)}&=& (\delta_{l0} +c_l^2)j_l(kR)^2 +\left(2 \delta_{l0} \sum_{n=1}^{\infty}\frac{a_n c_n}{2n+1} +2a_l c_l\right)kR j_l(kR) J_l'(kR) \nonumber\\
&&+a_l^2(kR)^2 (j_l'(kR))^2 +\delta_{l0}\sum_{n=1}^{\infty}\frac{a_n^2}{2n+1}(kR)^2 j_0(kR)j_0''(kR)~.
\label{unse_i2.eqn}
\end{eqnarray}
Using the following integration formulae,
\begin{eqnarray}
\int_0^\infty \mbox{d}k e^{-\epsilon k}(j_l(kR))^2 &=& \frac{\pi}{2(2l+1)R} + O(\epsilon)~, \label{unse_formula06.eqn} \\
\int_0^\infty \mbox{d}k e^{-\epsilon k} kR j_l(kR) j_l'(kR) &=&-\frac{\pi}{4(2l+1)R}+ O(\epsilon)~, \label{unse_formula07.eqn} \\
\int_0^\infty \mbox{d}k e^{-\epsilon k} (kR)^2 (j_l'(kR))^2 &=&\frac{1}{2\epsilon} -\frac{\pi l(l+1)}{2(2l+1)R} \nonumber\\
&&\hspace{1cm} +O(\epsilon, ~{\rm or}~\epsilon~\log \epsilon)~, \label{unse_formula08.eqn} \\
\int_0^\infty \mbox{d}k e^{-\epsilon k} (kR)^2j_0(kR)j_0''(kR) &=& -\frac{1}{2\epsilon} +\frac{\pi}{2R} \nonumber\\
&&\hspace{1cm} +O(\epsilon, ~{\rm or}~\epsilon~\log \epsilon)~,
\label{unse_formula09.eqn}
\end{eqnarray}
we obtain Eq.(\ref{uns_ec.eqn}) with Eq.(\ref{uns_hc.eqn}).

\section{DECAY RATE OF UNSCREENED F-BALL\label{uns_decay.sec}}

Unscreened charged F-balls are metastable and fragment through the quantum tunneling (see Sec.\ref{unscreened.sec} for the details). Here, we roughly estimate the rate of fragmentation, assuming a very simplified model of fragmentation after the ellipsoidal deformation from the spherical shape \footnote{The geometrical assumption of ellipsoidal shape enables us to estimate the upper bound on the lifetime of the F-ball.}.

We express the surface of the ellipsoidal F-ball as
\begin{equation}
\frac{x^2}{R^2}+\frac{y^2}{R^2}+\frac{z^2}{R^2\eta^2}=1~,
\label{unsd_surface.eqn}
\end{equation}
with $\eta$ being a constant which we call 'deformation parameter'. Note that $R$ depends on $\eta$ since it is determined as to minimize the energy of the F-ball. The surface is also expressed in the polar coordinates as
\begin{equation}
r=R_f(\theta)=\frac{\eta R}{\sqrt{\mbox{cos}^2\theta+\eta^2\mbox{sin}^2\theta}}~.
\label{unsd_r.eqn}
\end{equation}
The scalar field $\phi(r,\theta)$ should satisfy the Euler-Lagrange equation to be derived from Eq.(\ref{neu_lagrangian.eqn}),
\begin{equation}
\nabla^2\phi=\phi''+\frac{2}{r}\phi'+\frac{1}{r^2}\ddot{\phi}+\frac{1}{r^2\tan{\theta}}\dot{\phi}=\frac{\lambda}{2}\phi(\phi^2-v^2)~,
\label{unsd_eqM.eqn}
\end{equation}
with the symbols, 'dash' and 'dot', standing for $\partial/\partial r$ and $\partial/\partial \theta$, respectively. Here, the boundary conditions are imposed,
\begin{eqnarray}
\left\{
\begin{array}{l}
\phi(r\rightarrow\infty)=+v \\
\phi(r=R_f)=0 \\
\phi(r\rightarrow 0)\sim-v~~.
\end{array}
\right.
\label{unsd_bc.eqn}
\end{eqnarray}
(The last one becomes exact when $R_f$ tends to infinity (see Eq.(\ref{unsd_phi.eqn})).) Using the following variable,
\begin{equation}
\tilde{r}=r-R_f(\theta)~,
\label{unsd_rtilde.eqn}
\end{equation}
and keeping the terms of the leading order for $R_f\rightarrow \infty$, we can rewrite Eq.(\ref{unsd_eqM.eqn}) as
\begin{equation}
\frac{1}{\Theta^2}\frac{\partial^2\phi}{\partial \tilde{r}^2}=\frac{\lambda}{2}\phi(\phi^2-v^2)~,
\label{unsd_eqM2.eqn}
\end{equation}
to give the solution,
\begin{equation}
\phi=v\tanh{\left\{\Theta\frac{(r-R_f)}{\delta_n}\right\}}~,
\label{unsd_phi.eqn}
\end{equation}
with $\delta_n=2/\sqrt{\lambda}v$ and
\begin{eqnarray}
\Theta
&=&
\frac{1}
{\sqrt{1+\frac{1}{R_f^2}(\frac{\partial R_f}{\partial \theta})^2}}
\label{unsd_thetageneral.eqn} \\
&=&\frac{\cos{\theta}^2+\eta^2\sin{\theta}^2}{\sqrt{\cos{\theta}^2+\eta^4\sin{\theta}^2}}~. \label{unsd_theta.eqn}
\end{eqnarray}

We regard the deformation parameter $\eta$ as the time-dependent dynamical variable. Substituting $\phi$ into the Lagrangian density of Eq.(\ref{neu_lagrangian.eqn}) with the Fermi, the Coulomb and the volume effect added to it, and integrating it over the whole space, we get the Lagrangian,
\begin{equation}
L=\frac{1}{2}h(\eta)\dot{\eta}^2-W(\eta)~,
\label{unsd_lagrangian.eqn}
\end{equation}
where $h(\eta)$ and $W(\eta)$ are estimated as follows. First, $h(\eta)$ is expressed as
\begin{eqnarray}
h(\eta)&=&2\pi v^2\int\int\mbox{d}r\mbox{d}\theta~ r^2\sin{\theta}\left(\frac{\frac{\partial}{\partial\eta}\left\{\Theta\frac{(r-R_f)}{\delta_n}\right\}}{\mbox{cosh}^2\left\{\Theta\frac{(r-R_f)}{\delta_n}\right\}}\right)^2 \label{unsd_hdef.eqn} \\
&\simeq &4\pi\Sigma R^4\left\{-\frac{\eta^4}{R^2}\left(\frac{\partial R}{\partial \eta}\right)^2N_1+2\frac{\eta^3}{R}\left(\frac{\partial R}{\partial \eta}\right)N_2-\eta^2N_3\right\}~,
\label{unsd_h.eqn}
\end{eqnarray}
where
\begin{equation}
N_k=\int_0^1\mbox{d}x~x^{2(k-1)}\left\{(\eta^2-1)x^2-\eta^2\right\}^{-k}\left\{\eta^4-(\eta^4-1)x^2\right\}^{-1/2} \hspace{0.7cm}(k=1,2,3)~.
\label{unsd_nk.eqn}
\end{equation}
Since $1/\mbox{cosh}^4\left\{\Theta(r-R_f)/\delta_n\right\}$ in the integrand in Eq.(\ref{unsd_hdef.eqn}) is negligibly small except at around $r\sim R_f$, we approximate the quantities which are smooth at $r\sim R_f$  to those at $r=R_f$. Now let us estimate $W(\eta)=E_v+E_s+E_f+E_c$. We easily obtain the volume energy,
\begin{equation}
E_v=\frac{4\pi R^3\eta}{3}\epsilon~.
\label{unsd_ev.eqn}
\end{equation}
In order to estimate $E_s$, $E_f$ and $E_c$, we introduce new variables $\hat{\theta}$ and $\hat{\varphi}$ defined by
\begin{eqnarray}
\left\{
\begin{array}{l}
x=R\sin{\hat{\theta}}\cos{\hat{\varphi}} \\
y=R\sin{\hat{\theta}}\sin{\hat{\varphi}} \\
z=\eta R\cos{\hat{\theta}}~~.
\end{array}
\right.
\label{unsd_surface_redef.eqn}
\end{eqnarray}
The surface element is expressed as
\begin{eqnarray}
\mbox{d}S&=&R^2\sqrt{\mbox{cos}^2\hat{\theta}+\eta^2\mbox{sin}^2\hat{\theta}}\sin{\hat{\theta}}\mbox{d}\hat{\theta}\mbox{d}\hat{\varphi} \nonumber\\
&=&R^2\sqrt{k(u)}\mbox{d}u\mbox{d}\hat{\varphi}~,
\label{unsd_ds.eqn}
\end{eqnarray}
where $u=\cos{\hat{\theta}}$ and $k(u)=u^2+\eta^2(1-u^2)$.
From Eq.(\ref{unsd_ds.eqn}), we obtain the surface energy,
\begin{equation}
E_s=\Sigma \int \mbox{d}S=4\pi\Sigma R^2\int_{0}^{1}\mbox{d}u~\sqrt{k(u)}~.
\label{unsd_es.eqn}
\end{equation}
We express the fermion number density,
\begin{equation}
\sigma_f(u)=\frac{N_f}{4\pi R^2\sqrt{k(u)}}\left(1+\sum_{l=1}^{\infty}n_lP_l(u)\right)~,
\label{unsd_sigmaf.eqn}
\end{equation}
where $P_l(u)$ is the $l$'th order Legendre polynomial. It is easy to see that the above expression satisfies the condition for the fermion number $N_f$ to be conserved (see Eq.(\ref{neu_nftot.eqn})). From Eq.(\ref{unsd_ds.eqn}) and Eq.(\ref{unsd_sigmaf.eqn}), we obtain the fermi energy,
\begin{equation}
E_f=\frac{4\sqrt{\pi}}{3}\int\mbox{d}S~\sigma_f^{3/2}=\frac{2N_f^{3/2}}{3R}\int_{0}^{1}\mbox{d}u~\frac{\left(1+\sum_{l=1}^{\infty}n_lP_l(u)\right)^{3/2}}{k(u)^{1/4}}~.
\label{unsd_ef.eqn}
\end{equation}
Let us estimate the Coulomb energy,
\begin{equation}
E_c=\frac{e^2}{8\pi}\int\int\mbox{d}S\mbox{d}S'~ \frac{\sigma_f\sigma_f'}{\left| {\bf x}-{\bf x}' \right|}~.
\label{unsd_ecdef.eqn}
\end{equation}
Using the relation,
\begin{equation}
\frac{1}{4\pi|{\bf x}-{\bf x}'|}=\int\frac{\mbox{d}^3{\bf q}}{(2\pi)^3q^2}e^{i{\bf q}({\bf x}-{\bf x}')}=\int\frac{\mbox{d}^3\hat{\bf q}}{(2\pi)^3\hat{q}^2k(u_{\hat{q}})}e^{i\hat{\bf q}(\hat{\bf x}-\hat{\bf x}')}~,
\label{unsd_formula03.eqn}
\end{equation}
with
\begin{eqnarray}
\hat{\bf x}&=&(x_x, x_y, \frac{1}{\eta}x_z)=R(\sin{\hat{\theta}}\cos{\hat{\varphi}}, \sin{\hat{\theta}}\sin{\hat{\varphi}}, \cos{\hat{\theta}})~, \label{unsd_xhat.eqn} \\
\hat{\bf q}&=&(q_x, q_y, \eta q_z)~, \label{unsd_khat.eqn}
\end{eqnarray}
and $u_{\hat{q}}=\hat{q}_z/\hat{q}$, we get
\begin{equation}
E_c=\frac{e^2N_f^2\eta}{4\pi^2}\int_{0}^{\infty}\mbox{d}\hat{q}\int\frac{\mbox{d}\Omega_{\hat{q}}}{4\pi k(u_q)}\left| \int\frac{\mbox{d}\Omega_{\hat{x}}}{4\pi}~\left(1+\sum_{l=1}^{\infty}n_l P_l(u) \right)e^{i\hat{\bf q}\hat{\bf x}}\right|^2~.
\label{unsd_ec01.eqn}
\end{equation}
With the aid of 
\begin{equation}
\int_0^{\infty}\mbox{d}q~j_l(qR)j_{l'}(qR)=\frac{\pi}{2(2l+1)R}\delta_{ll'} \hspace{1cm} \mbox{for}~~l-l'=\mbox{even~ integer}~,
\label{unsd_jljl.eqn}
\end{equation}
the integration in Eq.(\ref{unsd_ec01.eqn}) can be carried out in the same way as in Appendix \ref{unse.sec}, to give
\begin{equation}
E_c=\frac{e^2N_f^2}{8\pi R}\int_{0}^{1}\mbox{d}u~\frac{\eta}{k(u)}\left(1+\sum_{l=1}^{\infty}\frac{n_l^2P_l(u)^2}{2l+1}\right)~.
\label{unsd_ec.eqn}
\end{equation}

\begin{figure}[htbp]
\includegraphics[height=8cm]{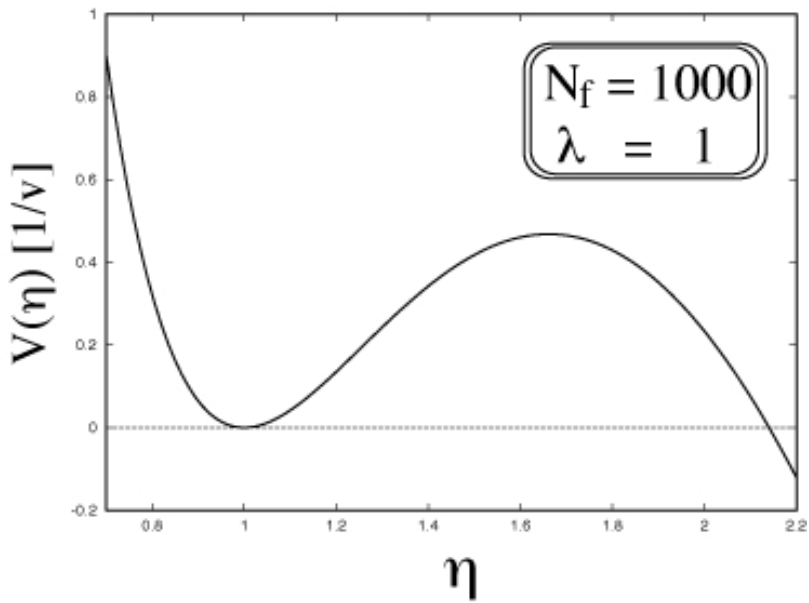}
\caption{The potential is displayed as a function of $\eta$ in case
the F-ball is not screened for the parameters shown in the graph.
There is an energy barrier, which is comparable to $v$.  The
distribution of the fermion is shown in Fig.\ref{uns_distr.fig}.
\label{uns_pot.fig}}
\end{figure}

We then minimize $W(\eta)$ with respect to $R$ and $n_l$. As expected from the perturbative analysis in Sec.\ref{unscreened.sec}, we verify numerically that $W(\eta)$ takes the local minimum value $W_{min}$ at $\eta=1$ in a certain range of the parameters (see Fig.\ref{uns_pot.fig} for the case, $\epsilon=0$, $\lambda=1$, and $N_f=1000$). 
\begin{figure}[htbp]
\includegraphics[height=8cm]{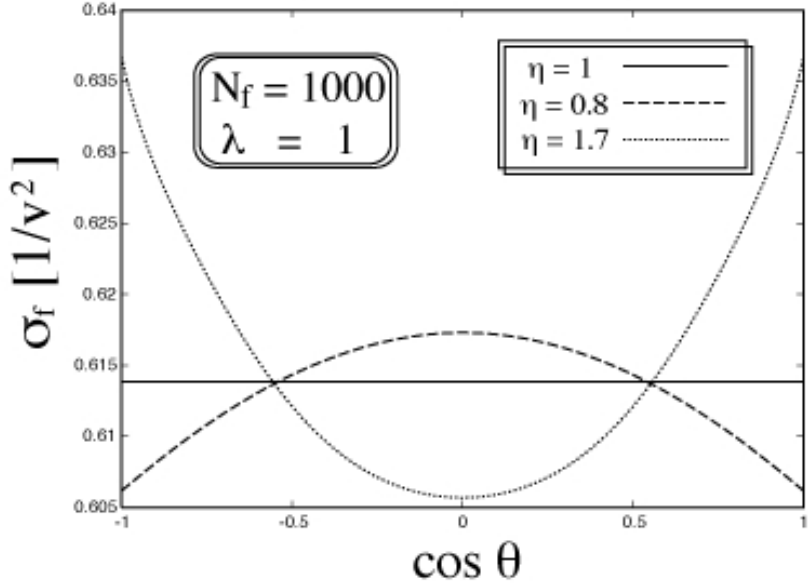}
\caption{The number density of the heavy fermion as a function of
$\theta$ for $\eta = 0.8,1,1.7$.  The density is constant for $\eta
=1$.  It gets large at the equator for $\eta =0.8$ and at poles for
$\eta =1.7$.  They are graphically displayed in Fig.\ref{cigar.fig}.
\label{uns_distr.fig}}
\end{figure}
\begin{figure}[htbp]
\includegraphics[height=6cm]{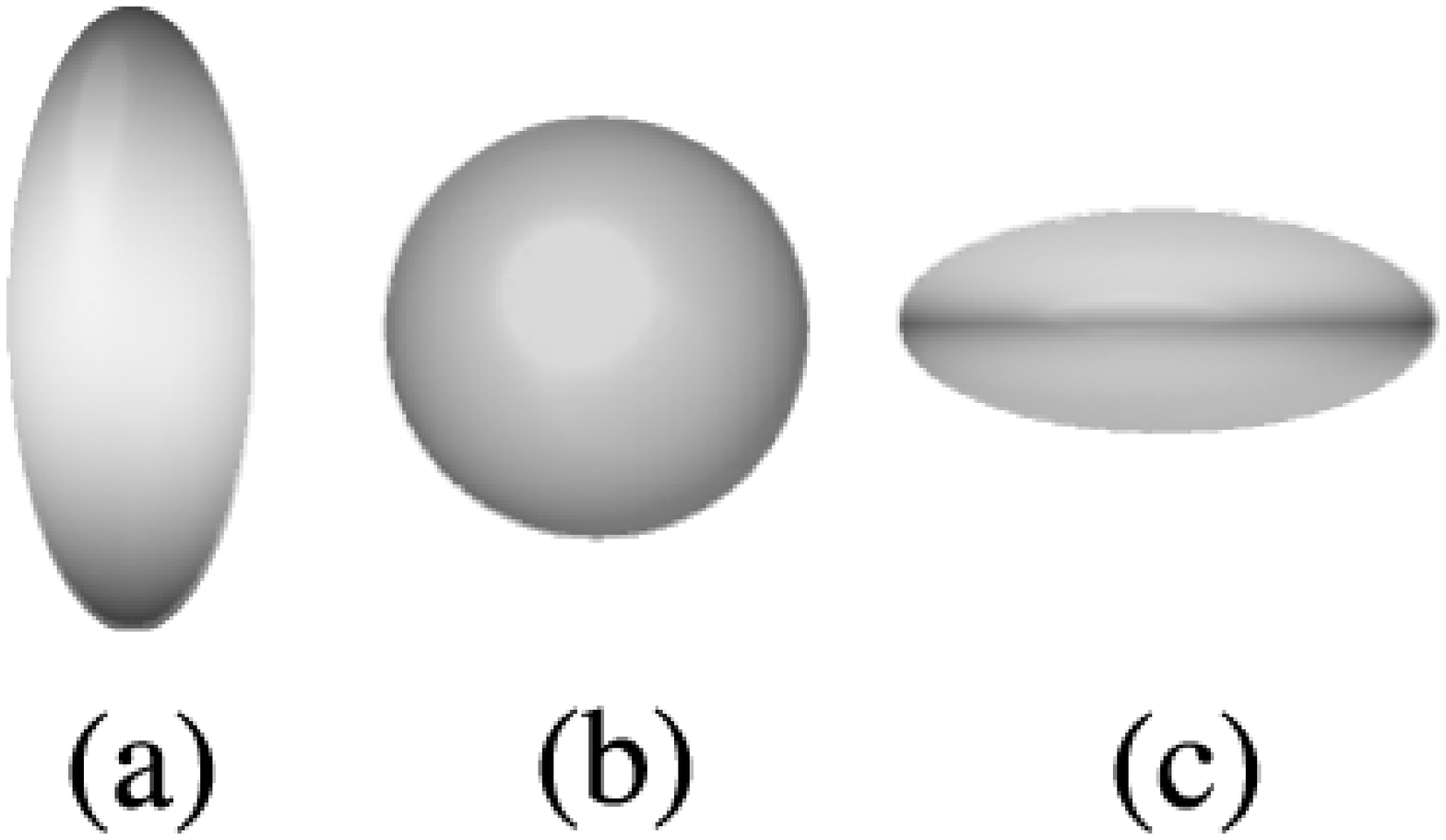}
\caption{The distribution of the heavy fermions in case the F-ball is
not screened.  (a) The F-ball becomes cigar-like shape for $\eta > 1$
and the fermions gather at poles due to the repulsive electric force.
(b) The F-ball is a sphere for $\eta =1$ and the distribution is
constant.  (c)The F-ball has the pancake-like shape for $\eta < 1$, and the fermions gather at the equator due to the repulsive force.
\label{cigar.fig}}
\end{figure}
In the following, we consider such an F-ball which has the local minimum value of $W$ at $\eta=1$, that is, is stable against the deformation from the spherical shape.

We finally proceed to estimate the decay rate of the metastable F-ball, using the Euclidean action method in Ref.\cite{Col1}. We express the Euclidean action as
\begin{equation}
{\cal A}=\int\mbox{d}t~\left\{\frac{1}{2}h(\eta)\dot{\eta}^2+V(\eta)\right\}~,
\label{unsd_actiondef.eqn}
\end{equation}
where $V(\eta)=W(\eta)-W_{min}$ (we note $V(1)=0$). 
Defining the following variable,
\begin{equation}
\xi=\int_{1}^{\eta}\mbox{d}\eta'\sqrt{h(\eta')}~,
\label{unsd_xidef.eqn}
\end{equation}
we express the action as
\begin{equation}
{\cal A}=\int\mbox{d}t~\left\{\frac{1}{2}\dot{\xi}^2+\overline{V}(\xi)\right\}~,
\label{unsd_action.eqn}
\end{equation}
where the potential $\overline{V}(\xi)=V(\eta(\xi))$ has a local minimum $\overline{V}=0$ at $\xi=0$. According to Ref.\cite{Col1}, the decay rate $\Gamma$ is estimated semi-classically,
\begin{equation}
\Gamma=\hbar\left|K\right|e^{-{\cal A}_0/\hbar}~,
\label{unsd_gamma.eqn}
\end{equation}
where ${\cal A}_0$ is the classical action of the bounce solution and $K$ is the prefactor, both of which are evaluated below. The bounce solution for $\xi$ should satisfy the following equation,
\begin{equation}
\frac{\delta {\cal A}}{\delta \xi}=-\ddot{\xi}+\overline{V}'(\xi)=0~,
\label{unsd_eqxi.eqn}
\end{equation}
with the boundary condition,
\begin{equation}
\xi \rightarrow 0 \hspace{2cm}\mbox{for}~~t\rightarrow \pm \infty~.
\label{unsd_bcxi.eqn}
\end{equation}
From Eq.(\ref{unsd_eqxi.eqn}), we get $\dot{\xi}^2=2\overline{V}(\xi)$ for the classical solution, and obtain the classical Euclidean action,
\begin{eqnarray}
{\cal A}_0&=&2\int_{0}^{\xi_c}\mbox{d}\xi~\sqrt{2\overline{V}(\xi)} \nonumber\\
&=&2\int_{1}^{\eta_c}\mbox{d}\eta~\sqrt{2h(\eta)V(\eta)}~,
\label{unsd_claction.eqn}
\end{eqnarray}
where $\eta_c$ ($\eta_c>1$) and $\xi_c$ ($\xi_c>0$) are defined as $V(\eta_c)=\overline{V}(\xi_c)=0$. The constant $K$ is defined as
\begin{equation}
K=\sqrt{\frac{{\cal A}_0}{2\pi\hbar}}\left\{\frac{\mbox{det}\left(-\frac{\partial^2}{\partial t^2}+\overline{V}''(0)\right)}{\mbox{det}'\left(-\frac{\partial^2}{\partial t^2}+\overline{V}''(\xi)\right)}\right\}^{1/2}~,
\label{unsd_kdef.eqn}
\end{equation}
where the symbol '$\mbox{det}'$' means the determinant in which the zero-mode contribution is removed \cite{Col1}. In our case, $K$ is calculated as
\begin{eqnarray}
K&=&\sqrt{\frac{\pi}{\hbar}}\left(\overline{V}''(0)\right)^{3/4}\xi_c\exp{\int_{0}^{\xi_c}\mbox{d}\xi\frac{\sqrt{\overline{V}''(0)}\xi-\sqrt{2\overline{V}(\xi)}}{\xi\sqrt{2\overline{V}(\xi)}}} \nonumber\\
&=&\sqrt{\frac{\pi}{\hbar}}\left(\frac{V''(1)}{h(1)}\right)^{3/4}\left(\int_{1}^{\eta_c}\mbox{d}\eta\sqrt{h(\eta)}\right) \exp{\int_{1}^{\eta_c}\mbox{d}\eta\sqrt{h(\eta)}} \nonumber\\
&&\hspace{1cm}\times\left\{\left(\frac{V''(1)}{2h(1)V(\eta)}\right)^{1/2}-\frac{1}{\int_{1}^{\eta}\mbox{d}\eta'\sqrt{h(\eta')}}\right\}~.
\label{unsd_k.eqn}
\end{eqnarray}

Taking parameters $v=10^5~\mbox{GeV}$, $N_f=1000$, and $\lambda=1$ for example, we obtain ${\cal A}_0\simeq 550$.  Thus, the lifetime of the unscreened charged F-ball is much longer than the age of the universe in this case. (If the production temperature of the F-ball $T_f$ is as high as $T_f\sim T_{ph}\sim v$, the F-ball would decay through the thermal fluctuation. We here consider the temperature $T_f\lesssim 0.1v\ll T_{ph}$ where the lifetime of the F-ball is found to be long enough to let it survive till present.)

\section{DECAY RATE OF SCREENED F-BALL\label{s_decay.sec}}

Here, we roughly estimate the decay rate of the screened F-ball, assuming the very simplified model, in which the surface has the
following shape,
\begin{eqnarray}
R_f
=
R_0 (1+a_2 P_2(\cos{\theta}))~,
\label{sd_r.eqn}
\end{eqnarray}
where $P_2$ is the 2nd order Legendre polynomial and $R_0$ and $a_2$ are arbitrary constants. This allows the F-ball to have a concave shape.  In the following we consider the F-ball energy as the function of the area $S$ and the parameter $a_2$. Then, the parameter $R_0$ is a function of $S$ and $a_2$.  We use the same technique as that in
Appendix \ref{uns_decay.sec}.  We here regard $a_2$ as the time-dependent dynamical variable instead of $\eta$ in Appendix \ref{uns_decay.sec}.  All we need to obtain
are $\tilde h(a_2)$ and $\tilde V(a_2)$ in the Lagrangian,
\begin{equation}
L=\frac{1}{2}\tilde h(a_2)\dot{a_2}^2-\tilde V(a_2)~.
\label{sd_lagrangian.eqn}
\end{equation}
We can easily obtain $\tilde h(a_2)$ by replacing $R_f$ in
Eqs.(\ref{unsd_thetageneral.eqn}) and (\ref{unsd_hdef.eqn}) with that defined by Eq.(\ref{sd_r.eqn}).

\begin{figure}[htbp]
\includegraphics[height=8cm]{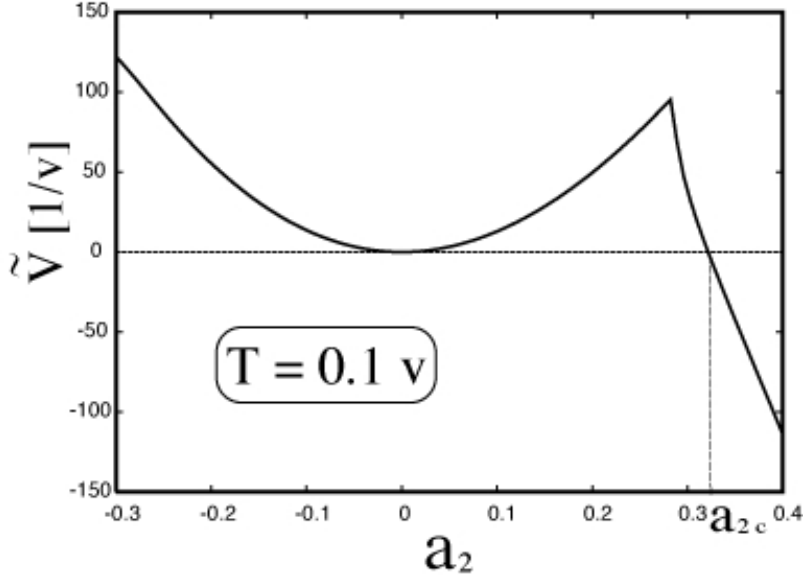}
\caption{The potential of the screened F-ball as a function of $a_2$.
The curvature at the origin is positive and the potential becomes negative
for $a_2$ larger than $a_{2c}$ at $T=0.1 v$. These facts mean that the
spherical F-ball ($a_2=0$) is metastable.  The energy barrier is much larger than the
temperature.  This tells us that the decay due to the thermal
fluctuation can be neglected.
\label{scr.fig}}
\end{figure}

\begin{figure}[htbp]
\includegraphics[width=10cm]{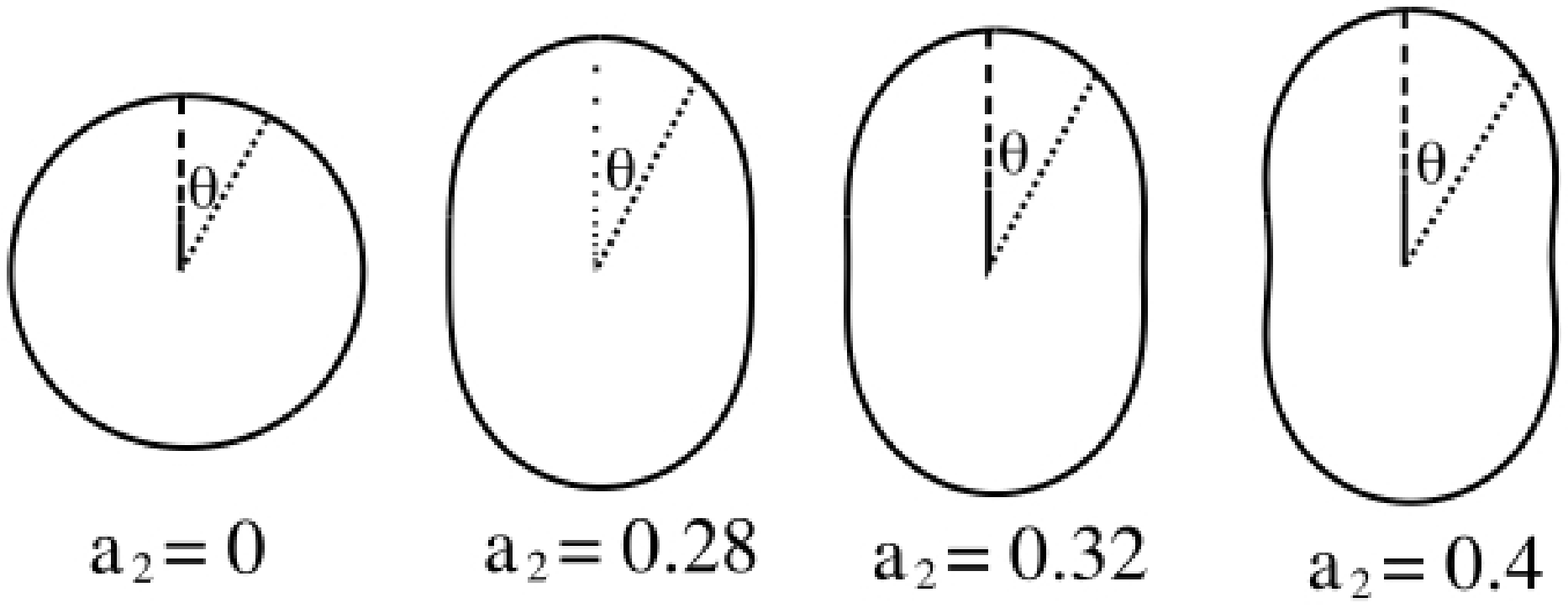}
\caption{The shapes of the F-ball for various values of the parameter $a_2$.  A geometrical cross section containing z-axis is shown.  The shape is a sphere
 for $a_2=0$.  One of the principal radii is divergent at the equator for
$a_2=2/7\sim 0.28$ in which the potential has the crest.  The shape
becomes slightly non-convex at the equator for $a_2=a_{2c}\sim 0.32$
and more narrow in the middle for $a_2 = 0.4$. 
\label{shape.fig}}
\end{figure}

The potential $\tilde V(a_2)$ can be obtained from
Eq.(\ref{scr_ftot2.eqn}) since $F_{tot}^{(0)}$ depends only on $S$ and not on the shape of the F-ball.  We calculate the following quantities using the
method  introduced in Appendix \ref{dv.sec},
\begin{eqnarray}
X_C
&\equiv &
\int\mbox{d}S~\frac{1}{\left| R_1R_2 \right|}~,\label{Xcdef.eqn} \\
X_D
&\equiv &
\int\mbox{d}S~\left(\frac{1}{R_1}-\frac{1}{R_2}\right)^2
\frac{\left| R_1R_2 \right|}{ R_1R_2}~.
\label{Xddef.eqn}
\end{eqnarray}
The surface is expressed as
\begin{eqnarray}
{\bf x}={\bf x}_f(\theta,\varphi)
=
R_f(\theta)
(\sin{\theta}\cos{\varphi},\sin{\theta}\sin{\varphi},\cos{\theta})~,
\end{eqnarray}
in the polar coordinates.  The first and the second forms are
\begin{eqnarray}
I
&=&
E(\theta,\phi) \mbox{d}\theta^2
+2F(\theta,\phi) \mbox{d}\theta \mbox{d}\phi
+G(\theta,\phi) \mbox{d}\phi^2~,\\
J
&=&
L(\theta,\phi) \mbox{d}\theta^2
+2M(\theta,\phi) \mbox{d}\theta \mbox{d}\phi
+N(\theta,\phi) \mbox{d}\phi^2~,
\end{eqnarray}
with
\begin{eqnarray}
E(\theta,\phi)
&=&
R_f^{2}+R_f^{'2}, \hspace{1cm} 
L(\theta,\phi)
=
\frac{1}{\sqrt{R_f^{2}+R_f^{'2}}}
(R_f^{''2}R_f^{2}-2R_f^{'2}-R_f^{2}), \nonumber \\
F(\theta,\phi)
&=&
0~
, \hspace{1cm} 
M(\theta,\phi)
=
0, \nonumber \\
G(\theta,\phi)
&=&
R_f^2 \sin^2{\theta}~
, \hspace{1cm} 
N(\theta,\phi)
=
\frac{R_f}{\sqrt{R_f^{2}+R_f^{'2}}}
\sin{\theta}(-R_f\sin{\theta}+R_f{\cos{\theta}})~,
\end{eqnarray}
where the symbol 'dash' denotes the derivative with respect to $\theta$. From these, the Gaussian curvature in Eq.(\ref{dv_gaussian.eqn}) and the
mean curvature in Eq.(\ref{dv_mean.eqn}) are expressed in terms of $R_f$.  We then obtain
\begin{eqnarray}
\frac{1}{R_1R_2}
&=&K=
\frac{
(R_f^{'}\sin{\theta}-R_f{\cos{\theta}})
(-R_f^{''2}R_f^{2}+2R_f^{'2}+R_f^{2})
}{
R_f \sin{\theta}
(R_f^{2}+R_f^{'2})^2
}~,\label{gaussian.eqn} \\
\left(\frac{1}{R_1}-\frac{1}{R_2}\right)^2
&=&
\frac{(EN-GL)^2}{E^2G^2}\nonumber\\
&=&
\frac{
\left\{
(R_f^{2}+R_f^{'2})(-R_f^{'}\sin{\theta}+R_f\cos{\theta})
+
(-R_f^{''2}R_f^{2}+2R_f^{'2}+R_f^{2})R_f\sin{\theta}
\right\}^2
}{
R_f^2\mbox{sin}^2\theta(R_f^{2}+R_f^{'2})^3
}~.\nonumber \\ \label{mean.eqn}
\end{eqnarray}
From these expressions, $X_C$ and $X_D$ are written as
\begin{eqnarray}
X_C
&=&
4\pi
\int_0^{\frac{\pi}{2}}\mbox{d}\theta
\frac{
\left|
(R_f^{'}\sin{\theta}-R_f{\cos{\theta}})
(-R_f^{''2}R_f^{2}+2R_f^{'2}+R_f^{2})
\right|
}{
(R_f^{2}+R_f^{'2})^{\frac{3}{2}}
}~,\label{Xc.eqn} \\
X_D
&=&
4\pi
\int_0^{\frac{\pi}{2}}\mbox{d}\theta
\frac{
\left\{
(R_f^{2}+R_f^{'2})(-R_f^{'}\sin{\theta}+R_f{\cos{\theta}})
+
(-R_f^{''2}R_f^{2}+2R_f^{'2}+R_f^{2})R_f\sin{\theta}
\right\}^2
}{
R_f \sin{\theta}
(R_f^{2}+R_f^{'2})^{\frac{5}{2}}
}
\frac{|K|}{K}~.\nonumber\\ \label{Xd.eqn}
\end{eqnarray}
We note that these quantities do not explicitly depend on the
parameter $R_0$ or $S$ and only depend on $a_2$.   Using the free energy,
\begin{eqnarray}
F^{(2)}_{tot}(a_2)
=
(C_n+C_{sc})X_C+(D_n+D_{sc})X_D~,
\label{f2tot.eqn}
\end{eqnarray}
we evaluate the potential,
\begin{eqnarray}
\tilde V(a_2)
=
F^{(2)}_{tot}(a_2)-F^{(2)}_{tot}(0)~.
\end{eqnarray}
We show the potential as a function of $a_2$ in Fig.\ref{scr.fig} for the case $T=0.1v$. 
The cusp in Fig.\ref{scr.fig} arises because of the non-analytic feature of $X_C$ and $X_D$.
At the top of the cusp, the Gaussian curvature becomes zero at equator
and turns to be negative for larger $a_2$.  This is illustrated in Fig.\ref{shape.fig}.
One can observe that the curvature at the origin
is positive and the potential becomes negative for $a_2$ larger than
$a_{2c}$. These facts mean that the spherical F-ball is metastable as discussed
in Sec.\ref{screened.sec}.  Because the energy barrier is much larger
than the temperature, the decay of the F-ball due to thermal
fluctuation can be neglected.  We next consider the decay due to
the tunneling effect.

We consider the classical Euclidean action, 
\begin{eqnarray}
\tilde{\cal  A}_0
=
2\int_{0}^{a_{2c}}\mbox{d}a_2~
\sqrt{2\tilde h(a_2)\tilde V(a_2)}~.
\label{sd_claction.eqn}
\end{eqnarray}
We can apply Eq.(\ref{unsd_hdef.eqn}) to obtain $\tilde{h}(a_2)$ and see  its magnitude is $O(\Sigma R^4)$.  We find that the barrier height is $O(100v)$
and $a_{2c}$ is $O(1)$ from Fig.\ref{scr.fig}.  From these, the classical
Euclidean action is roughly estimated as $\tilde{\cal A}_0\gtrsim
N_f$.  Since the number of the heavy fermion is large, $N_f > 10^{19}$, from
Eqs.(\ref{con_nf.eqn}) and (\ref{con_v.eqn}), the classical
Euclidean action is extremely large and the screened F-ball is
effectively stable.  Although we use the very simple deformation model and take the special case of $T=0.1v$, we think that the result of very long lifetime is rather general for the temperature smaller than $v$ since the result is only due to the macroscopic property of the F-ball.

\begin{acknowledgments}
We are grateful to Makoto Sakamoto, Masahiro Kawasaki, Shigeki Matsumoto, Masahide Yamaguchi and Masamune Oguri for the useful comments. We would like to thank Kojin Takeda and Hisaki Hatanaka for the helpful conversations. One of the authors (K.O.) is a research fellow of JSPS(No.4834).
\end{acknowledgments}


\end{document}